%% file: BFJR_GG.tex
\documentclass[fleqn, usenatbib, twocolumn]{mnras}

\usepackage{newtxtext,newtxmath}
\usepackage[T1]{fontenc}
\usepackage{orcidlink}

\DeclareRobustCommand{\VAN}[3]{#2}
\let\VANthebibliography\thebibliography
\def\thebibliography{\DeclareRobustCommand{\VAN}[3]{##3}\VANthebibliography}

\usepackage{graphicx}	
\usepackage{amsmath}	
\usepackage[para,online]{threeparttable}

\newcommand{\percent}{\ensuremath{\%}}
\newcommand{\Msun}{M_\odot}
\newcommand{\Mgas}{M_\mathrm{gas}}
\newcommand{\Mbar}{M_\mathrm{bar}}
\newcommand{\Mstar}{M_\mathrm{star}}
\newcommand{\gbar}{\mathrm{g}_\mathrm{bar}}
\newcommand{\gobs}{\mathrm{g}_\mathrm{obs}}
\newcommand{\gdag}{\mathrm{g}_\mathrm{\dagger}}
\newcommand{\gddag}{\mathrm{g}_\mathrm{\ddagger}}

\defcitealias{Angus08}{A08}
\defcitealias{MK11}{MK11}

\title[BFJR in galaxy groups]{Examining Baryonic Faber-Jackson Relation in Galaxy Groups}

\author[Pradyumna \& Tian]{
Pradyumna Sadhu\,\orcidlink{0000-0002-3947-6239}$^{1}$\thanks{E-mail: psadh003@ucr.edu} and
Yong Tian\,\orcidlink{0000-0001-9962-1816}$^{2}$\thanks{E-mail: yongtian@astro.ncu.edu.tw}
\\
$^{1}$Department of Physics and Astronomy, University of California, Riverside, CA 92507, USA\\
$^{2}$Institute of Astronomy, National Central University, Taoyuan 320317, Taiwan \\}

\date{Accepted 2024 January 29. Received 2024 January 8; in original form 2023 May 7}

\pubyear{2024}

\begin{document}
\label{firstpage}
\pagerange{\pageref{firstpage}--\pageref{lastpage}}
\maketitle

\begin{abstract}
We investigate the Baryonic Faber-Jackson Relation (BFJR), examining the correlation between baryonic mass and velocity dispersion in galaxy groups and clusters. Originally analysed in elliptical galaxies, the BFJR is derivable from the empirical Radial Acceleration Relation (RAR) and MOdified Newtonian Dynamics (MOND), both showcasing a characteristic acceleration scale $\mathrm{g}_\mathrm{\dagger}=1.2\times10^{-10}\,\mathrm{m}\,\mathrm{s}^{-2}$. Recent interpretations within MOND suggest that galaxy group dynamics can be explained solely by baryonic mass, hinting at a BFJR with $g_{\dagger}$ in these systems. To explore this BFJR, we combined X-ray and optical measurements for six galaxy clusters and 13 groups, calculating baryonic masses by combining X-ray gas and stellar mass estimates. Simultaneously, we computed spatially resolved velocity dispersion profiles from membership data using the biweight scale in radial bins. Our results indicate that the BFJR in galaxy groups, using total velocity dispersion, aligns with MOND predictions. Conversely, galaxy clusters exhibit a parallel BFJR with a larger acceleration scale. Analysis using tail velocity dispersion in galaxy groups shows a leftward deviation from the BFJR. Additionally, stacked velocity dispersion profiles reveal two distinct types: declining and flat, based on two parallel BFJRs. The declining profile, if not due to the anisotropy parameters or the incomplete membership, suggests a deviation from standard dark matter density profiles. We further identify three galaxy groups with unusually low dark matter fractions.
\end{abstract}




\section{Introduction}

Kinematic scaling relations provide insights into the intricacies of the ``missing mass'' problem and baryonic physics in virialized systems like galaxies, galaxy groups, and clusters. 
In these systems, the ``missing mass'' was repeatedly observed in kinematics, dynamics, and even gravitational lensing, well-known as the Dark Matter (DM) problem when assuming a hypothetical non-baryonic particle as the solution. The kinematic scaling relations can provide important insights into constraints on DM and baryons \citep[e.g.,][]{Famaey12, Capelo2012, Desmond2015, Desmond2017a, WT2018}, baryonic physics like feedback \citep[e.g.,][]{Voit2005, Dutton2009, Andreon2010, Fabian2012, Pasini2020}, and evolution of these systems in general~\citep{Mo1998, vandenBosch2000, VonDerLinden2007}. Further, reproducing the slope, intercept, and scatters of these relations provides serious tests to various theories \citep{McGaugh2012, Courteau2007, Papastergis2016, Desmond2017b, Tian21a, Tian21b}.

Arguably, the most extensively studied kinematic scaling relation in galaxies is the Baryonic Tully-Fisher relation \citep[BTFR,][]{McGaugh2000, Verheijen2001, McGaugh2011, McGaugh2012, Lelli2019}. 
In spiral galaxies, the BTFR is a tight empirical correlation between the baryonic mass $\Mbar$ and the flat rotational velocity $V_\mathrm{f}$ with the form $\Mbar\propto\,V_\mathrm{f}^4$.
Compared to its predecessor Tully-Fisher relation~\citep{Tully1977} between luminosity and flat rotational velocity, the BTFR demonstrates a smaller scatter while covering a wider dynamical range of masses~\citep{McGaugh2011, McGaugh2012, Lelli2019}. The intercept of the BTFR implies a characteristic acceleration scale $g_\dagger = 1.2 \times 10^{-10}\,\mathrm{m}\,\mathrm{s}^{-2}$~\citep{McGaugh2018}. The small intrinsic tightness and the inherent acceleration scale pose a serious challenge to the popular $\Lambda$ Cold Dark Matter ($\Lambda$CDM) model~\citep{Famaey12, Desmond2015, Lelli2016}. Besides, \cite{Lelli2017} showed that the slope and intercept of the BTFR in galaxies could be deduced from the dynamical relation of spiral galaxies known as Radial Acceleration Relation \citep[RAR,][]{McGaugh2016}. 

The RAR is a tight dynamical scaling relation between (i) observed centripetal acceleration, $\gobs(r)$ obtained from the rotation curve at a given radius $r$, and (ii) the acceleration due to baryons alone, given by $\gbar(r)\equiv\, - \nabla \Phi_\mathrm{bar}(r)$, where $\Phi_\mathrm{bar}(r)$ represents the potential due to the baryonic matter at radius $r$.
Importantly, $\gobs$ and $\gbar$ are independently measured from the rotation curve and observed baryons, respectively.
The low acceleration limit ($\gbar\ll\gdag$) of the RAR can be expressed as~\citep{McGaugh2016}
\begin{equation}
    \gobs \simeq \sqrt{\gbar\gdag}\,.
    \label{eq:RARla}
\end{equation}
This low acceleration limit of the RAR implies the BTFR~\citep{Lelli2017} in spiral galaxies.
Coincidentally, the best-fit value of $\gdag$ of the RAR~\citep{McGaugh2016, Li2018, Ghari2019} is identical to the one measured in the BTFR. Given its surprisingly small scatter, the RAR has been seen as a strong tool to constrain feedback within $\Lambda$CDM \citep{Paranjape2021}, and also to test other theories~\citep[e.g.,][]{Ren2019, Berezhiani2018}. 
More recently, \cite{Li2022} found difficulty in explaining the observed RAR in the $\Lambda$CDM model even after employing physical processes such as halo compression. Coincidentally, in elliptical galaxies, the RAR implies the famous Faber-Jackson Relation~\citep[FJR,][]{Faber1976} with an assumption of a constant mass-to-light ratio \citep{Lelli2017}.

The FJR, between luminosity $L$ and the central velocity dispersion $\sigma$, is a counterpart of TFR for pressure-supported systems like elliptical galaxies \citep{Faber1976}. The FJR is a relation of the form $L \propto \sigma^{[3:5.3]}$. For this FJR, the exponent was observed to be dependent on the luminosity \citep{Choi2007, Desroches2007, Nigoche-Netro2010}, and whether the galaxy is in cluster environments \citep{VonDerLinden2007, Desroches2007}.

The Baryonic Faber-Jackson Relation~\citep[BFJR,][]{Sanders2010} is the baryonic version of the FJR that links the baryonic mass $\Mbar$ with the line-of-sight (los) flat velocity dispersion $\sigma_{\mathrm{los}}$.
This relation has, however, not been studied widely in elliptical galaxies. While the FJR was argued to be a projection of the fundamental plane in elliptical galaxies \citep{Dressler1987, Djorgovski1987}, later,  \cite{Sanders2010} showed the BFJR to subsume systems in a greater range of masses than the fundamental plane. 

The BFJR, the BTFR, and the RAR were predicted by an effective dynamical law about four decades ago, Modified Newtonian Dynamics or Milgrom Dynamics~\citep[MOND,][]{Milgrom1983}. 
MOND can naturally explain the RAR as well as its low acceleration limit implications 
 such as the BFJR and the BTFR.
Additionally, MOND has been built on an inherent acceleration scale $a_0$, which is identical to the measured $\gdag$.
MOND successfully made several unique predictions and explained many ``missing mass'' problems from parsec to kiloparsec scales~\citep{Famaey12, McGaugh20, Milgrom20, BZ22}. 
For example, MOND in pressure-supported systems and isothermal spheres 
 resembles the BFJR, and is given as~\citep{Milgrom1984, Milgrom2014}
\begin{equation}
    \centering
    \Mbar\,G\,a_0 = \frac{81}{4}\,\sigma_{\mathrm{los}}^4\,.
    \label{eq:BFJR_MOND}
\end{equation}
However, on megaparsec scales, MOND has difficulties explaining the ``missing mass'' 
 in galaxy clusters with observed baryons alone~\citep[e.g., see][]{Sanders1999, Sanders2003, Angus08, Famaey12}.

 In galaxy clusters, the missing residual mass within MOND gave rise to several possibilities 
 in the literature~\citep[e.g., see section 6.6.4 in][]{Famaey12},
 which includes (i) falsification of MOND;
(ii) missing baryons or non-baryonic dark matter \citep[for e.g., see][]{Milgrom2008, Angus08};
(iii) introduction of a second scale;
(iv) the requirement of the additional field in relativistic MOND.
Inspecting the possibilities of MOND requires investigating the RAR and the BFJR on cluster scales.
Even if the residual missing mass falsifies MOND, the success of MOND on galaxy scales still needs to be explained.

Inspired by MOND, the kinematic scaling relations in galaxy clusters have been explored for three decades. \cite{Sanders1994} was the first to indirectly examine the BFJR in galaxy clusters using the correlation between X-ray gas mass $\Mgas$ and temperature $T$, $\Mgas\propto\,T^2$, with temperature as a substitute for velocity dispersion.
Nevertheless, the acceleration scale implied by the BFJR in clusters was unclear because of the scatter and insufficient measurements.
Afterward, \citet{Zhang2011} combined X-ray gas and optical measurements to obtain the FJR as $L_{\mathrm{gas}}\propto\sigma^{4.46\pm0.23}$, which is not exact BFJR due 
 to incomplete baryonic mass and flat velocity dispersion estimates. 
Particularly, more recent studies~\citep{Tian21a, Tian21b} have discovered a tight parallel BFJR on the BCG-cluster scales, albeit with a ten-times larger acceleration scale $\gddag=(1.7\pm0.7)\times10^{-9}$\,m\,s$^{-2}$. 
In the studies by~\cite{Tian21a, Tian21b}, 29 galaxy clusters and 53 BCGs are found to have a consistent kinematic scaling relation from the centre of the cluster (marked by the BCG) to the outermost membership. This region, from the BCG to the furthest members of the cluster, is what we define as the "BCG-cluster scale".
To be distinguished from the BFJR, \citet{Tian21a, Tian21b} addressed this tight kinematic scaling relation as the Mass-Velocity Dispersion Relation (MVDR), given by
\begin{equation}\label{eq:MVDR}
    \centering
    \log\left(\frac{\Mbar}{\Msun}\right)=4.1^{+0.1}_{-0.1}\log\left(\frac{\sigma_{\mathrm los}}{\mathrm{km}\,\mathrm{s}^{-1}}\right)+1.6^{+0.3}_{-0.3}\,,
\end{equation}
with a tiny error from the lognormal intrinsic scatter of $10^{+2}_{-1}\percent$. 
Coincidentally, the MVDR is in agreement with the RAR on BCG-cluster scales, $\gobs\simeq\sqrt{\gbar\gddag}$, computed from lensing studies in \cite{Tian20}. 
On the other hand, other works also found a similar offset of the RAR in galaxy clusters ~\citep{Chan2020, Pradyumna2021, Pradyumna2021b, Ettori22, Tam23, Liu2023, Li2023}. However, these studies have found the RAR in clusters to either be an ambiguous correlation or to have a larger scatter, as compared to the RAR in galaxies.

While a tenfold increase in the acceleration scale $\gddag$ on galaxy clusters might contradict the universal acceleration constant $a_0$ posited by MOND, certain relativistic MOND theories could potentially account for 
$\gddag$~\citep{ZF12, HZ17, SZ21}. For instance, \citet{ZF12, HZ17} proposed an Extended MOND (EMOND) that boosts the acceleration scale in regions of high potential, thereby amplifying gravitational effects in galaxy clusters. Conversely, \citet{SZ21} developed a relativistic gravitational theory that not only replicates MOND in weak-field quasistatic situations, but also explains observed cosmic microwave background and matter power spectra on linear cosmological scales. These theories could potentially provide alternative gravity estimations in galaxy clusters, diverging from traditional MOND predictions.

Because galaxy groups are relatively diffuse with baryonic masses similar to the large galaxies but size comparable to galaxy clusters, these provide a unique testbed for MOND. The study of empirical scaling relations is paramount in galaxy groups because these are more abundant than the galaxy clusters \citep{Lovisari2021}. However, groups pose difficulties for both optical and X-ray measurements because of low richness, associated projection effects, and shallow gravitational potential. Recently, \citet{Sohn2019} examined the FJR in galaxy groups by identifying spectroscopic members of X-ray systems to get $L_{\mathrm{gas}}\propto\,\sigma^{4.7\pm0.7}$. Another relation associated with the BFJR is  $M_\mathrm{gas}-T$, which has been studied in \cite{Bahar2022}. The $T-\sigma$ relation for the richness range considered in the current work can be found in \cite{Wilson2016}.
Combining both these relations gives $M_\mathrm{gas} 
\propto \sigma^{3.06 \pm 0.51}$ as the first estimate of the BFJR in groups. However, a direct empirical relation between the baryonic mass and flat velocity dispersion tail still needs to be explored.

The galaxy groups cannot be considered scaled-down versions of more massive galaxy clusters owing to different stellar and gas fractions \citep[e.g.,][]{Andreon2010, Giodini09, Gonzalez2007}, and therefore different levels of feedback. 
Furthermore, \cite{Lovisari2015} have shown low mass clusters to deviate 
 from the scaling relations observed by the higher mass galaxy clusters.
However, the difference between galaxy groups and clusters has not been explicit in the literature \citep{Lovisari2021}. 
Thus, it is worthwhile to check whether the groups and clusters lie on the same kinematic scaling relations.
 
While the BFJR has been studied on galaxy and galaxy cluster scales, it has not been investigated in galaxy groups with spatially resolved velocity dispersion profiles including X-ray gas and stellar masses. 
The BFJR in galaxies implies an acceleration scale $\gdag$~\citep{Sanders2010}, while, the MVDR on the BCG-cluster scale indicates a larger acceleration scale $\gddag$~\citep{Tian21a, Tian21b}.
In addition, \cite{Milgrom2018, Milgrom2019} has conformed MOND in galaxy groups by examining various MONDian relations, which further implies the existence of the BFJR.
On the other hand, \cite{Hernandez2012, Durazo2017} indicate deviations of the velocity dispersion profile from the Newtonian expectations when the acceleration of the system falls below $a_0$. Since accelerations are expected to fall below $a_0$ in galaxy groups, investigating the velocity dispersion profile and the empirical BFJR in galaxy groups is essential to fill the gap in the literature.

This paper is organised as follows.
In Sec.~\ref{sec:data and methods}, we detail the samples of galaxy clusters and groups employed in the current study and the analysis performed. The BFJR obtained for our sample, and our other results are described in Sec.~\ref{sec:results}. The implications of our results are discussed in Sec.~\ref{sec:discussion}, and we summarise our study in Sec.~\ref{sec:conclusion}. We assume flat $\Lambda$CDM cosmology with $H_0 = 70\mathrm{\, km\, s^{-1}}$, $\Omega_\mathrm{m} = 0.3$ and $\Omega_{\Lambda}=0.7$ for deducing the angular diameter distances from redshifts.

\section{Data and Methods}
\label{sec:data and methods}

Investigating the BFJR in galaxy groups and clusters requires both the flat tail velocity dispersions and the baryonic masses.
Consequently, the velocity dispersion profile has to be computed by identifying member galaxies from optical surveys. 
On the other hand, the baryonic mass of a group/cluster $\Mbar$ includes the estimation of X-ray gas mass $\Mgas$ and the stellar mass of member galaxies $\Mstar$.
However, collecting a complete catalogue with both X-ray and optical measurements was not viable, especially for galaxy groups. 
Therefore, we studied an X-ray sample and an optical sample separately by obtaining their counterparts from the literature.
Namely, we adopted (i) an X-ray sample of 26 relaxed galaxy groups and clusters from \citet[][hereafter \citetalias{Angus08}]{Angus08}; and (ii) an optical sample of 395 galaxy groups identified by \citet[][hereafter \citetalias{MK11}]{MK11}.

For completing the X-ray sample of 26 groups and clusters in \citetalias{Angus08}, 
 we gathered the membership information and the optical \textit{K-}band luminosities from SIMBAD \citep{simbad}.
The member galaxies are carefully identified by excluding the repeated and uncertain members for each group/cluster in our sample. Due to statistical validity, only groups/clusters with membership greater than 25 are selected. Consequently, only six galaxy clusters remain with both X-ray gas and optical measurements. The basic quantities and references of our subsamples are listed in Table~\ref{tab:table1}.

To be consistent with the analysis of the \citetalias{Angus08} sample, 
we adopted the same selection criteria for the member galaxies of \citetalias{MK11} galaxy groups, and 
 assembled X-ray counterparts from the literature accordingly.
\citetalias{MK11} provided membership and \textit{K-}band luminosities for 395 galaxy groups\footnote{\url{https://relay.sao.ru/hq/dim/groups/}}, which were identified with a group finding algorithm in HyperLEDA and NED databases. 
On selecting groups having more than 25 member galaxies, the number of groups dramatically dropped 
 to 24 groups.
Among them, only 13 groups have redshift measurements available in the NED database\footnote{\url{http://ned.ipac.caltech.edu/}}, which we employ in our study. 
Finally, we complete X-ray counterparts for nine out of 13 groups using \cite{Babyk18} and \cite{yamasaki2009}.
However, we do not exclude the four groups without X-ray measurements since $\Mgas$ may not dominate over $\Mstar$ in galaxy groups \citep{Giodini09}. 
Table~\ref{tab:table1} lists the main properties of 13 \citetalias{MK11} groups studied in our work.

\input{Table1}

\subsection{Velocity Dispersion Profiles}\label{sec:vdp}

Following \cite{Tian21a}, we compute and study the los velocity dispersion profiles, where member galaxies are assumed to be tracers of the gravitational potential. Firstly, we determined the geometric centres for each of the 19 systems. Centroids and mean los velocities were readily provided by \citetalias{MK11} for their groups\footnote{For consistency within this work, the velocities provided in Local Group rest frame by MK11 were converted into the heliocentric frame following \cite{Karachentsev1996}.}.
On the other hand, we considered BCGs as the geometric centres in \citetalias{Angus08} systems, according to \cite{Zabludoff1998} and \cite{Lin04}. Relative to these centres, velocities of member galaxies are calculated and are binned radially. The total number of bins for a given group/cluster was chosen based on the membership such that there are at least 13 member galaxies in each bin. Finally, velocity dispersion was calculated in each of the bins. 

\begin{figure*}
\centering
\includegraphics[width=0.8\paperwidth]{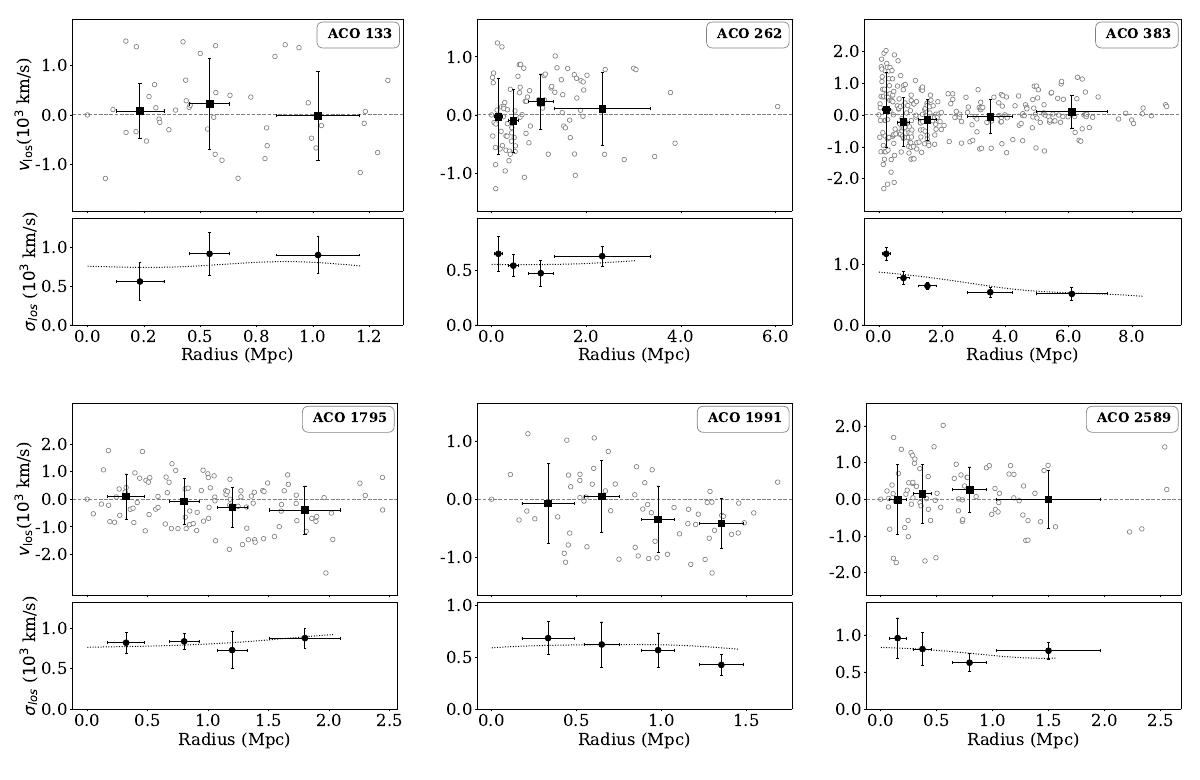}
\caption{
Velocity dispersion profiles of six galaxy clusters in \protect\citetalias{Angus08}: ACO133, ACO262, ACO383, ACO1795, ACO1991, and ACO2589; 
\textit{Upper panel of subplot:} The unfilled smaller grey circles show the los velocities of member galaxies relative to the geometric centre ($V_{\mathrm{los}}$) against the projected radii. 
The black-filled squares are binned los velocities relative to the centre binned using the biweight technique.
\textit{Lower panel of subplot:} The black filled circles with the error bars indicate the los velocity dispersions $\sigma_{\mathrm{los}}$ estimated in individual bins using the biweight estimator. 
The dotted line indicates a continuous velocity dispersion profile obtained following \protect\cite{Annie2009}.
}
\label{fig:vd_profiles1}
\end{figure*}

Given only a handful of points in each bin, we adopted the biweight method \citep{Beers1990} to calculate the velocity dispersion, since we expect deviations from the Gaussian distribution. For relatively small datasets, the biweight estimator is more robust to outliers than the standard deviation. In our samples, discrepancies up to $20\percent$ were observed in the velocity dispersion estimates using a biweight estimator and standard deviation. Thus, our concern about implementing the robust method is justified. In addition, to determine the uncertainties, we performed bootstrapping, see Figures~\ref{fig:vd_profiles1}, \ref{fig:vd_profiles2} and \ref{fig:vd_profiles3}.

For investigating the BFJR, we considered velocity dispersion in the last bin (equivalently, outer radii) as the representative of the velocity dispersion profile.
We labelled this as the flat velocity dispersion to be consistent with \cite{Tian21a, Tian21b}.
Contrary to the traditional studies like \cite{Nigoche-Netro2010} where central velocity dispersion is employed in the study of the Faber-Jackson relation, we use flat velocity dispersion since a lower scatter is expected in BFJR within the MOND paradigm. 
\cite{Sanders2010} showed the BFJR to have different intercepts within MOND for Newtonian and deep-MOND limits. Thus, by choosing flat velocity dispersion in outer radii where the deep-MOND limit holds, a lower scatter is expected in BFJR within MOND.
In addition, \cite{Durazo2017} find the velocity dispersion profile to asymptote, with a BFJR for the asymptotic velocity dispersion.
Consequently, we considered velocity dispersion in the last bin for our study of the BFJR since it is expected to be the closest to the flat velocity dispersion.

The flat velocity dispersions are listed in Table~\ref{tab:table1} under $\sigma_\mathrm{los,f}$. It must be noted that the velocity dispersion computed in the last bin is presumably larger than the true flat velocity dispersion for galaxy groups, since most of the galaxy groups present a declining velocity dispersion profile. Besides, we list the accelerations calculated using $\sigma_\mathrm{los,f}^2/R$ in Table~\ref{tab:table1}, where $R$ is the projected radius of the last bin. The accelerations for both \citetalias{Angus08} and \citetalias{MK11} samples are lower than the MOND acceleration scale $a_0$, as anticipated.

Besides binning data to obtain a discrete velocity dispersion profile, we computed the continuous profiles by implementing a moving window function $w_i$ suggested in \cite{Bergond2006, Annie2009}. The $w_i$ at a radius $R$ is given by
\begin{equation}
    w_i(R) = \frac{1}{\sigma_R} \exp\left[-\frac{(R-R_i)^2}{2 \sigma_R^2}\right]\,,
\end{equation}
where $i_\mathrm{th}$ member is at projected radius $R_i$ and $\sigma_R$ is the window width determining the contribution of distant points. As it was pointed out by \cite{Annie2009}, too small $\sigma_R$ captures unnecessary features, whereas too large $\sigma_R$ results in ignorance of necessary features. In this work, after some trials, $\sigma_R$ was chosen as 20$\percent$ of the projected radius of the most distant galaxy member. With this window function, we can obtain the velocity dispersion profile as:
\begin{equation}
    \sigma_p(R) = \sqrt{\frac{\sum_i v_\mathrm{los, i}^2 w_i(R) }{\sum_i w_i(R)}}\,,
\end{equation}
where $v_\mathrm{los, i}$ represents the los velocity of $i_\mathrm{th}$ member galaxy, and summations are performed over all the member galaxies. In the lower panel of Figures~\ref{fig:vd_profiles1}, \ref{fig:vd_profiles2} and \ref{fig:vd_profiles3}, the dotted line indicates the continuous profile computed by this method, which is seen to be consistent with the biweight method.

To compare our work with \cite{Milgrom2018, Milgrom2019}, we also measure the total velocity dispersion for each of the systems considered in this study. The total velocity dispersion (represented as $\sigma_\mathrm{tot}$) is calculated evaluating the biweight scale for velocities of all the members of the groups/clusters considered in this work. As discussed later in this paper, since most of the \citetalias{MK11} groups have a declining velocity dispersion profile, the total velocity dispersion is greater than the flat velocity dispersion for these systems. However, both total and flat velocity dispersions are similar for most of the \citetalias{Angus08} clusters (except ACO 383), since the clusters display flatter velocity dispersion profiles.

\subsection{The Baryonic Mass}

In this work, masses of X-ray emitting gas were taken from literature for both \citetalias{Angus08} and \citetalias{MK11} samples. \cite{Vikhlinin2006} and \cite{Zappacosta2006} provided the masses of X-ray gas within $R_{500}$, estimated using \textit{Chandra} and \textit{ROSAT} observations for the six clusters from \citetalias{Angus08}. Here, $R_{500}$ is the radius within which, the mean density of the cluster equals 500 times the critical density of the universe. For \citetalias{MK11} groups, the gas masses were collected from \cite{Babyk18} and \cite{yamasaki2009} for nine of the 13 groups. From Table~\ref{tab:table1}, gas masses for the \citetalias{Angus08} sample are seen to be in a range $(1.8 - 72.4) \times 10^{13} M_\odot$, whereas for \citetalias{MK11} sample, the gas masses are much lower, and in a range $(0.13 - 640) \times 10^{9}M_\odot$.

We employed $K$-band luminosities to estimate the stellar mass because these are considered robust proxies of the stellar mass irrespective of star formation history \citep{Kauffmann98}. 
Specifically, to estimate the stellar mass of member galaxies $(M_\mathrm{star})$ from $K$-band absolute magnitude ($M_K$), the relation
\begin{equation}
    \log_{10}M_\mathrm{star} = 10.58 - 0.44(M_K + 23),
    \label{eq:mstar}
\end{equation}
from \cite{Capellari2013} was used. The stellar masses of the member galaxies are then combined to obtain the total stellar mass of a group or cluster.
This relation was obtained by performing linear regression to stellar masses determined from dynamical modelling against \textit{K-}band luminosities for ATLAS$^\mathrm{3D}$ sample of early-type galaxies. A limitation of this relation is that it is not calibrated at $M_\mathrm{star} > 10^{11.5}\;M_\odot$. 
However, there are only ten of the member galaxies with stellar mass above $10^{11.5}M_\odot$, one galaxy above $10^{11.75}M_\odot$ and no galaxies above $10^{12} M_\odot$. Thus, we do not expect large deviations from the stellar mass estimated using this relation.

We obtained the stellar masses for both of our samples from \textit{K-}band luminosities, using Equation~(\ref{eq:mstar}). The $K$-band luminosities for member galaxies of \citetalias{Angus08} clusters were obtained from SIMBAD. On the other hand, $K$-band luminosities for member galaxies of MK11 groups were provided in \cite{MK11} itself. 
We calculated the average stellar mass-to-light ratio in the $K$-band,
 $\langle\Upsilon_\mathrm{K}\rangle=0.92\pm0.22$, 
 which is in agreement with the previous measurement of $\langle\Upsilon_\mathrm{K}\rangle=0.95\pm0.29$ obtained by \cite{Bell2003, Williams2009}.
The known sources of uncertainty in the stellar mass estimated from the Equation~(\ref{eq:mstar}) are: (i) the intrinsic scatter of Equation~(\ref{eq:mstar}), (ii) uncertainty in distance, (iii) uncertainty in \textit{K}-band luminosity, and (iv) uncertainty in slope and intercept of Equation~(\ref{eq:mstar}). Since the contribution from (i) was much larger than others, we considered the intrinsic scatter of $14\percent$ in Equation~(\ref{eq:mstar}) \citep{Capellari2013} as the uncertainty in stellar masses estimated. Table~\ref{tab:table1} lists the stellar masses along with their uncertainties for the 19 systems. Clearly, the stellar mass contributes by more than 50$\percent$ in most \citetalias{MK11} galaxy groups as opposed to a maximum of 29$\percent$ in \citetalias{Angus08} clusters.

The X-ray gas mass and the stellar masses are summed for baryonic mass (see Table~\ref{tab:table1}). 
Evidently, the baryonic masses of galaxy clusters and groups differ by at least an order of magnitude. 
Furthermore, it can be noted that that logarithms of stellar mass and the baryonic mass are comparable for \citetalias{MK11} groups, and therefore, it remains justified to not exclude the four groups (NGC3031, NGC3311, NGC3992, and ESO507-025 groups) without X-ray gas mass measurements.

In our samples, the stellar fraction among the baryonic components demonstrates distinct features for galaxy clusters and groups, respectively.
With the definition of the stellar fraction $f_{\mathrm{star}}\equiv\Mstar/\Mbar$,
 two different fractions are found for cluster samples in \citetalias{Angus08} 
 and group samples in \citetalias{MK11} respectively.
From Table~\ref{tab:table1}, the median $\langle\,f_{\mathrm{star}}\rangle=0.14$ in \citetalias{Angus08} ranges from $0.03$ to $0.35$,
 whereas the median $\langle\,f_{\mathrm{star}}\rangle=0.92$ in \citetalias{MK11} ranges from $0.73$ to $0.99$.
In summary, X-ray gas constitutes the dominant component of baryonic mass in galaxy clusters, 
 while it remains a minor component in galaxy groups. It must also be noted that the gas masses from \citetalias{Angus08} may not represent the total gas content of galaxy clusters, since these are the masses inside $R_{500}$, and the best-fit gas mass profiles are seen to be increasing for the clusters at $R_{500}$ in \cite{Vikhlinin2006}. This further calls for the study of cluster samples with measurement till the cluster outskirts, such as the XCOP \citep{Eckert2017}.

\begin{figure*}
\centering
\includegraphics[width=0.8\paperwidth]{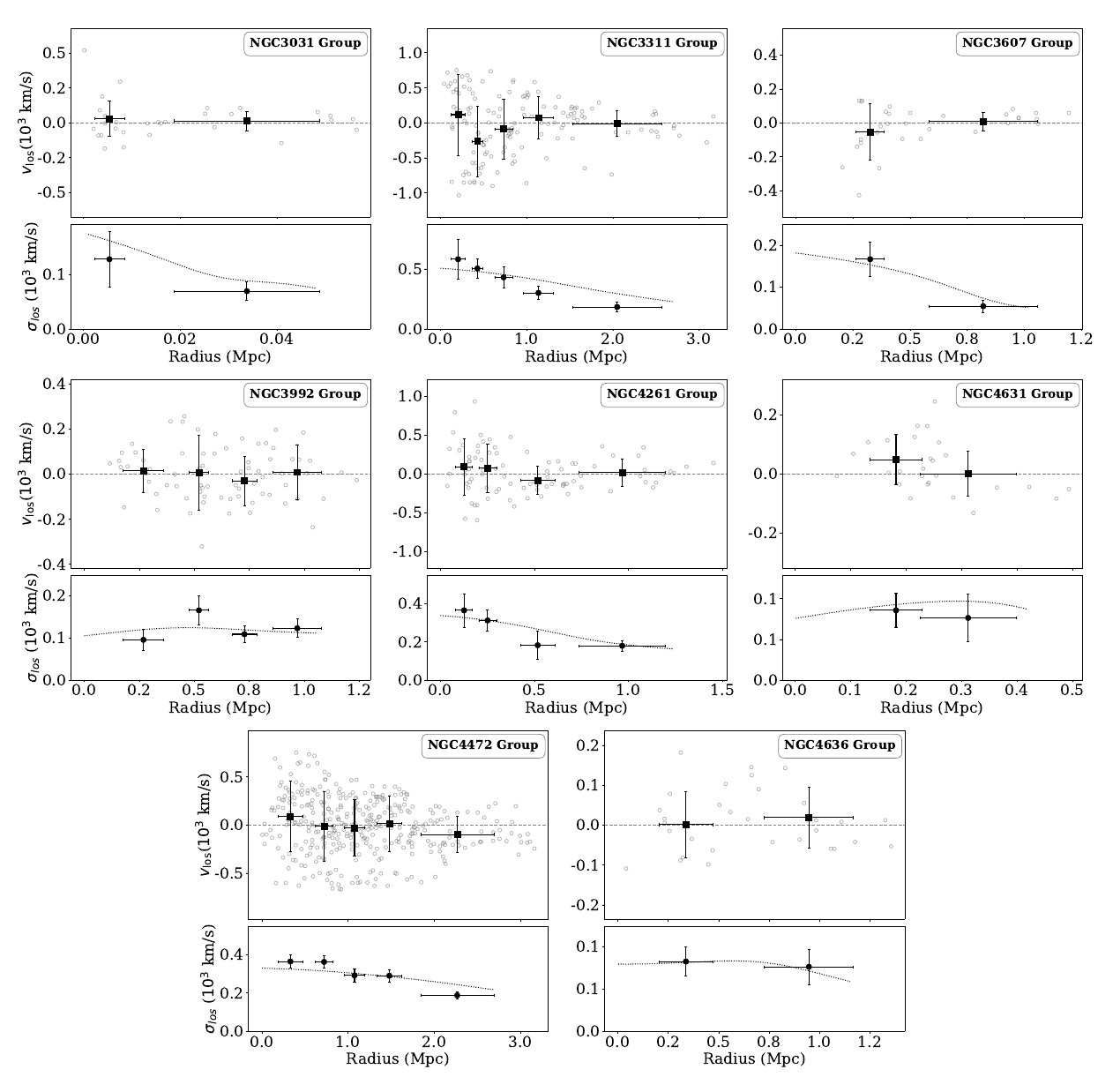}
\caption{
Velocity dispersion profiles of galaxy groups in \protect\citetalias{MK11}: NGC3031, NGC3311, NGC3607,NGC3992, NGC4261, NGC4631, NGC4472 and NGC4636. 
Symbols and lines follow the same conventions as in Figure~\ref{fig:vd_profiles1}.
}
\label{fig:vd_profiles2}
\end{figure*}

\begin{figure*}
\begin{center}
\includegraphics[width=0.8\paperwidth]{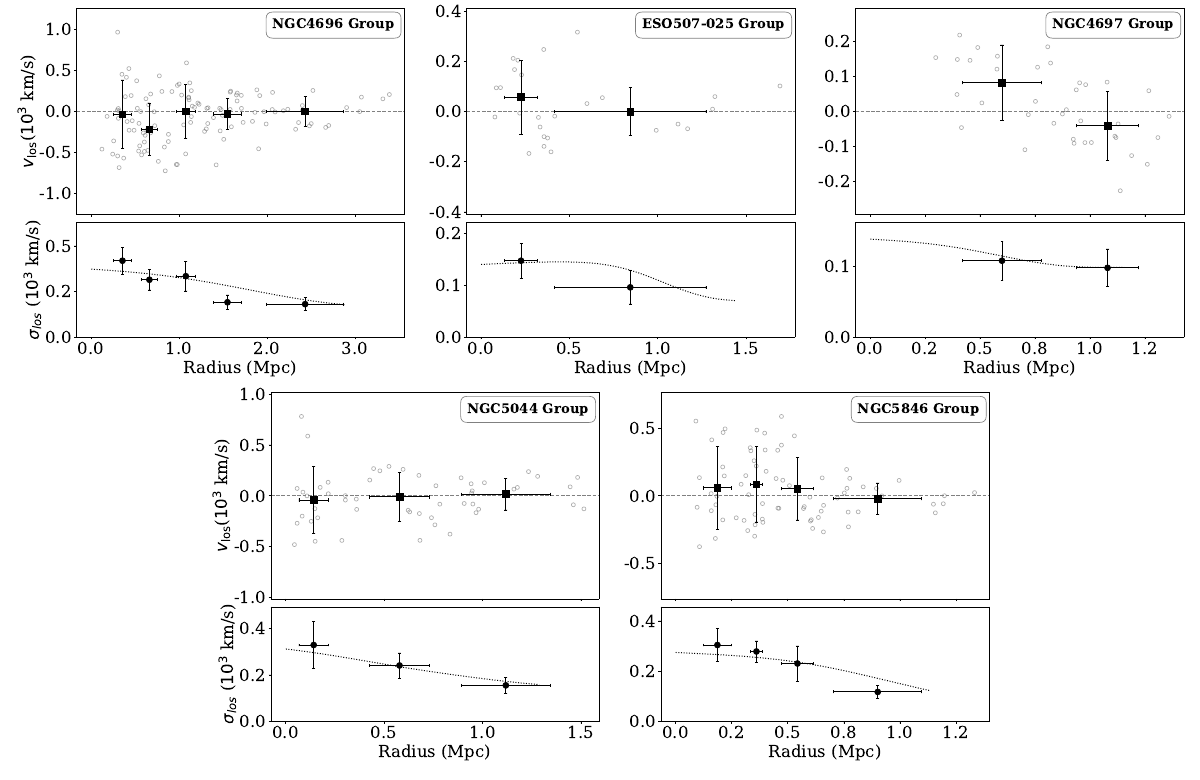}    
\end{center}

\contcaption{
Velocity dispersion profiles of galaxy groups in \protect\citetalias{MK11}: NGC4696, NGC4697, ESO507-025, NGC5044, and NGC5846.
Symbols and lines follow the same conventions as in Figure~\ref{fig:vd_profiles1}.
    }
\label{fig:vd_profiles3}
\end{figure*}

\section{Results}\label{sec:results}

We explore the BFJR, the corresponding acceleration scales, and the spatially resolved kinematic profiles in 19 systems. 
Our main goal is to investigate the BFJR in galaxy groups by studying the correlation between the baryonic masses and their flat los velocity dispersions. We additionally study the correlation between the baryonic mass and total velocity dispersion, since BFJR has traditionally been studied with the total velocity dispersion.
Because the BFJR can be implied by the RAR and MOND, we compare our results against the BFJR predicted by MOND.
Besides, we examine our results against other similar works~\citep{Milgrom2018, Milgrom2019, Tian21a, Tian21b}. Finally, we carefully identify two distinct features of the velocity dispersion profiles in our samples: the flat and the declining profiles. 
The BFJR, MVDR, and two distinct profiles were seen to correspond to their acceleration scales.

\subsection{Baryonic Faber-Jackson Relation}

\begin{figure*}
\centering
\includegraphics[width= 2\columnwidth]{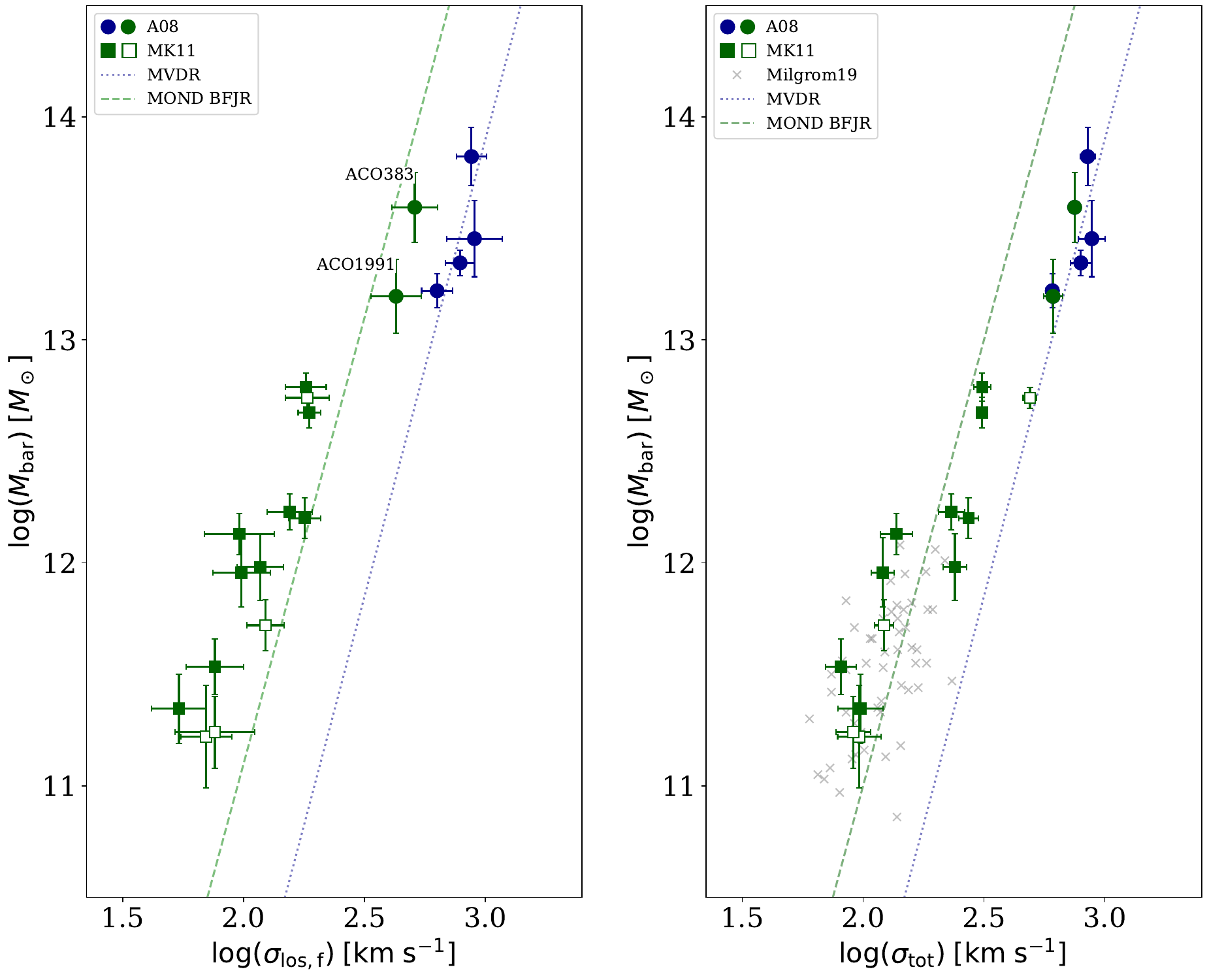}
\caption{
The Baryonic Faber-Jackson Relation for our 19 systems. Green and blue filled circles represent 
 galaxy clusters in \citetalias{Angus08}; Green filled squares are galaxy groups in \citetalias{MK11},
while the unfilled squares indicate groups with underestimated baryonic mass due to a lack of X-ray gas measurements. The green dashed line represents the BFJR given by Equation~(\ref{eq:BFJR})
The blue dotted line demonstrates the MVDR, Equation~(\ref{eq:MVDR}), discovered on BCG-cluster scales by \protect\cite{Tian21a}. The stellar masses are obtained using Equation (\protect\ref{eq:mstar}) in this figure. \textit{Left panel}: The BFJR when flat los velocity dispersion ($\sigma_\mathrm{los, f}$) is considered. Subsample-A (Green circles, filled and unfilled squares) is identified by their proximity to the BFJR as compared to MVDR in this panel.
On the other hand, Subsample-B (blue circles) is identified as the set of systems consistent with the MVDR. 
\textit{Right panel}: The BFJR when the total velocity dispersion ($\sigma_\mathrm{tot}$) is considered. The grey crosses are 56 galaxy groups studied in \protect\cite{Milgrom2019}. Most of the points can be seen to shift to the right as compared to the left panel because total velocity dispersion is greater than the flat velocity dispersion for systems with declining velocity dispersion profiles.}

\label{fig:MVDR}
\end{figure*}

Firstly, we plot flat los velocity dispersion $\sigma_{\mathrm{los,f}}$ against baryonic mass $\Mbar$ in logarithmic scale for six clusters of \citetalias{Angus08} and 13 groups of \citetalias{MK11}, see left panel of Figure~\ref{fig:MVDR}. 
Notably, most galaxy groups present a linear correlation in the logarithm plane, whereas the six massive galaxy clusters (marked in green and blue circles) deviate to the right.

To understand the linear correlation in our samples, the BFJR is needed as a reference in the logarithmic plane. Because the BFJR is predicted by MOND as Equation~(\ref{eq:BFJR_MOND}), we recast this BFJR in logarithmic form as $\log(\Mbar)=4\log(\sigma_{\mathrm{los}})+\log(81/4Ga_{0})\,$. When compared with our samples, we calculate the MOND prediction with physical units as
\begin{equation}\label{eq:BFJR}
 \log\left(\frac{\Mbar}{\Msun}\right)=4\log\left(\frac{\sigma_\mathrm {los}}{\mathrm{km}\,\mathrm{s}^{-1}}\right)+3.1\,,
\end{equation}
which is demonstrated as the green dashed line in Figure~\ref{fig:MVDR}.
All the 13 \citetalias{MK11} groups deviate from the BFJR implied within MOND, and thus are inconsistent with the simple prediction. Nevertheless, the inconsistency of these galaxy groups with the BFJR from MOND cannot be taken as a rejection of MOND, since Equation (\ref{eq:BFJR_MOND}) was derived for isothermal spheres.
However, the galaxy groups in this work are not isothermal spheres 
 since these do not abide by a constant velocity dispersion profile.
On the other hand, the other four \citetalias{Angus08} clusters (ACO133, ACO262, ACO1795, and ACO2589) deviate from Equation~(\ref{eq:BFJR}),  but are  consistent with the MVDR~\citep{Tian21a, Tian21b}. 
Notably, the MVDR has a smaller intercept, implying a larger acceleration scale $\gddag$, rather than $a_{0}$ or $\gdag$ of the BFJR. In contrast, two of the \citetalias{Angus08} clusters (ACO383 and ACO1991) lie in between BFJR and MVDR, however, closer to the former.

When total velocity dispersion is considered, most of the systems shift to the right on the baryonic mass - velocity dispersion plane, see the right panel of Figure~\ref{fig:MVDR}. Conventionally, BFJR has been studied using total velocity dispersion \citep[for e.g., ][]{Sanders2010, Nigoche-Netro2010, Milgrom2018, Milgrom2019}. We find the vertical scatter of \citetalias{MK11} groups about the MOND BFJR to decrease from 0.66 dex to 0.48 dex when flat velocity dispersion is replaced with the total velocity dispersion. Moreover, the points in the right panel seem to start from the left of the MOND BFJR, and gradually shift rightwards with increasing baryonic mass. 
 
Figure~\ref{fig:MVDR} also highlights the sample of 56 low-richness galaxy groups examined in \cite{Milgrom2019} in the right panel. 
It is essential to note that \cite{Milgrom2019} focuses on galaxy groups from \cite{MK11} consisting of 15 or fewer members, since the contribution of gas can be assumed to be negligible in these galaxy groups. We compare these points with our results utilising the total velocity dispersion, since they employ total velocity dispersion for studying the BFJR. 
The sample analysed in \citet{Milgrom2019} generally aligns with the set of 13 \citetalias{MK11} groups studied in this work.

\subsection{Two Distinct Velocity Dispersion Profiles}

Our 19 samples are divided into two subsamples based on their proximity to the two parallel kinematic scaling relations (BFJR and MVDR):\\
 A. 13 groups in \citetalias{MK11} marked with the green squares 
    plus two clusters (ACO383 and ACO1991) in \citetalias{Angus08} marked with the green circles in Figure~\ref{fig:MVDR} are closer to the BFJR in the left panel, hereafter subsample-A; \\
 B. four clusters (ACO133, ACO262, ACO1795, and ACO2589) in \citetalias{Angus08} 
 marked with blue circles in Figure~\ref{fig:MVDR} are closer to the MVDR in the left panel, hereafter subsample-B.

\begin{figure}
\centering
\includegraphics[width=1\columnwidth]{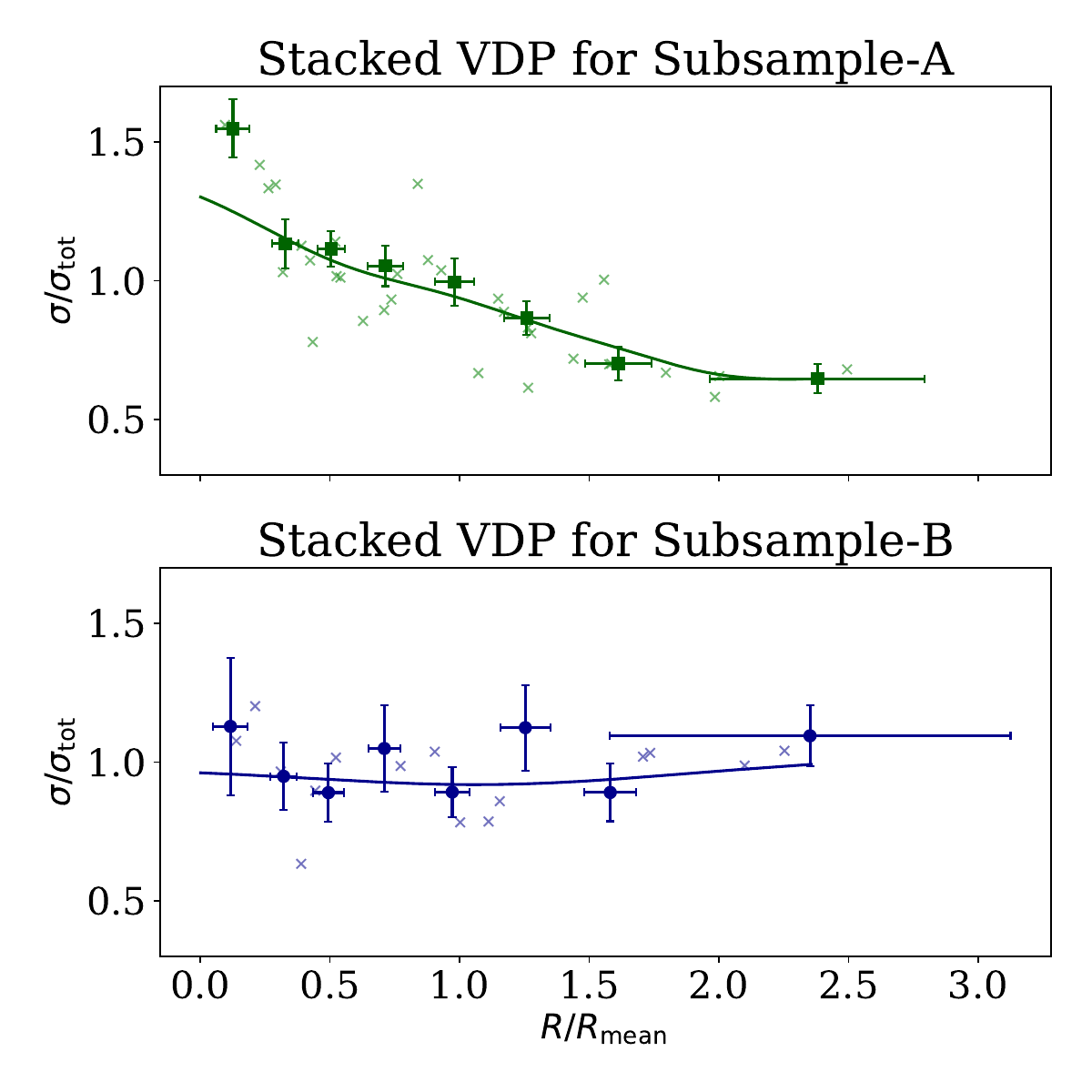}
\caption{Stacked velocity dispersion profiles after appropriate scaling for 13 groups and two clusters close to BFJR are shown in green (Subsample-A), and four clusters close to MVDR are shown in blue (Subsample-B). The stacked profile is obtained as detailed in Sec.~\ref{sec:morphology}. 
The smaller crosses in the background are scaled velocity dispersions in each bin of individual groups and clusters.
The bigger solid points are biweight scales obtained in respective bins for the combined sample, and the solid line is the continuous velocity dispersion profile obtained following the procedure laid out in Section~\ref{sec:vdp}}. 
\label{fig:svdp}
\end{figure}

We stacked the normalised velocity dispersion profiles to compare the two subsamples which correspond to the parallel kinematic scaling relations. 
For each group/cluster, the binned los velocities are normalised by the total velocity dispersion,\ $\sigma_\mathrm{tot}$. Further, the projected radii of these bins are normalised by mean distance to member galaxies, $R_\mathrm{mean}$\footnote{While choosing to scale the projected radii with a mean distance of member galaxies is arbitrary, we do not observe any noticeable difference in the final results when scaled by median or 75$^\mathrm{th}$ percentile radii.}. These scaled points were then combined for each of the subsamples, and are shown as crosses in Figure~\ref{fig:svdp}. To now compute the stacked velocity dispersion profile for these points, the analysis of Section~\ref{sec:vdp} is repeated.
Notably, the two subsamples show distinct features:
 (i) the declining profile, the upper panel (subsample A) in Figure~\ref{fig:svdp};
 and (ii) the flat profile, the lower panel (subsample B) in Figure~\ref{fig:svdp}.

We indeed discover two distinct features with the stacked profiles, based on their position in the baryonic mass-velocity dispersion plane.
The subsample-A presents a declining profile as shown in the upper panel in Figure~\ref{fig:svdp}, whereas the subsample-B demonstrates a flat profile in the lower panel in Figure~\ref{fig:svdp}.
Our findings for a declining profile are consistent with \cite{Li12}, who also reported similar declining profiles for galaxy groups within the stellar mass range examined in this study.
Our results are also supported by \cite{Tian21a, Tian21b}, where most BCGs and galaxy clusters have illustrated nearly flat velocity dispersion profiles. Remarkably, the existence of two distinct profiles raises an issue regarding a possible kinematic difference in profile between galaxy groups and clusters, assuming this is not due to incomplete memberships.

\section{Discussion}
\label{sec:discussion}

By studying the kinematic profile of our systems, we present the velocity dispersion profiles of 19 galaxy groups/clusters, and the correlation between the flat/total velocity dispersions and the baryonic mass. Incidentally, the two stacked profiles correspond to flat and declining profiles in our systems.

The additional detailed analyses include validation of the equilibrium state, inference for the $\Lambda$CDM model, comparison with MOND, and implications of the acceleration scale, which are examined in the following subsections. We study the deviation from the equilibrium state in our samples to justify the validity of this work for further implications and comparisons. Because MOND predicted a specific BFJR with a characteristic acceleration scale $a_0$, it is important to compare our results with the MOND prediction.

\subsection{Dynamical State of the Sample}\label{sec:morphology}

Because the dynamical equilibrium of the system may affect the validity of our study,
it is important to check dynamical state of the sample.
Although we inspect on the galaxy clusters (\citetalias{Angus08}) determined to be relaxed from X-ray measurements, 
 we still double-check them with the kinematics of member galaxies from the optical counterpart.
Especially, the dynamical state of \citetalias{MK11} groups can only be understood from member galaxies because systematic X-ray studies are absent for this sample.

The gaussianity of the los velocities of member galaxies is known to provide information about the dynamical state of groups and clusters \citep[e.g., see ][]{Annie2009, Costa18}. 
Thus, following \cite{Annie2009}, we applied the Anderson-Darling test to the los velocities of member galaxies for a given group/cluster to evaluate its gaussianity. 
A system is considered non-gaussian only if Gaussianity is rejected at more than 95\% significance by the Anderson-Darling test. 
Consistent with \cite{Angus08}, we determined all the six clusters taken from their work to be relaxed. However, from the \citetalias{MK11} sample, we determined five of the groups -- NGC3031, NGC3311, NGC3607, NGC4472, NGC5846 -- to be dynamically unstable. 
Given the dynamic equilibrium within \citetalias{Angus08} clusters, we may assume the velocity dispersion profile computed from member galaxies to closely match that obtained from the gas temperature profile, though the baryonic content of the clusters is dominated by the gas.

Contrary to \cite{Annie2009}, we noticed no correlation between the dynamical state of the system and its velocity dispersion profile on studying the stacked profiles.
Since \cite{Annie2009} mentions decreasing and flat velocity dispersion profiles for Gaussian and non-gaussian groups, respectively, we stack the Gaussian and non-gaussian groups of \citetalias{MK11} separately.
However, we found a decreasing profile for both Gaussian and non-gaussian groups without any manifest difference. 

\subsection{The Implications for the Dark Matter Problem}

The declining velocity dispersion profiles in our systems may be attributed to three causes: 
\begin{enumerate}
    \item  unconventional dark matter density profiles; 
    \item  the anisotropy parameter;
    \item  the incomplete membership.
\end{enumerate}
In the Jeans equation of pressure-supported systems~\citep{BT08}, the velocity dispersion profile is influenced by both the density profile and the anisotropy parameter. For example, the observed declining velocity dispersion profiles in elliptical galaxies are indicative of a lack of dark matter~\citep{Romanowsky03, MS03, TK16}. However, there is a degeneracy between the density profile and the anisotropy parameter in the Jeans equation, which should be resolved through Bayesian analysis~\citep{Li2023}. In the context of isothermal dark matter, a flat velocity dispersion profile would typically be expected, rather than a declining one. Even in the case of a Navarro, Frenk, and White profile \citep[NFW; ][]{Navarro1997}, the los velocity dispersion profile observed inside the virial radius is not expected to be strongly declining \citep{Lokas2001} for low concentration halos. Therefore, our samples exhibit an atypical density profile, deviating from conventional expectations.
Furthermore, the completeness of membership can also affect the velocity dispersion profiles, particularly by missing the high-velocity member galaxies at the outer radii in our samples. However, should this be the primary cause, our samples would then exhibit significant deviations from the Gaussian distribution.

 Estimates of dynamical mass indicate three galaxy groups that have little or no dark matter. By working out the dynamical mass using the formula $M_\mathrm{dyn} \approx r\sigma^2_\mathrm{los,f}/G$, 
 we address the dynamical mass of \citetalias{MK11} galaxy groups at their last binned data point. The median fraction of $M_\mathrm{dyn}/\Mbar$ is 11.35 in \citetalias{MK11} galaxy groups, which is consistent with previous studies of galaxy groups \citep[for e.g., ][]{Giodini09, Dai10, Gonzalez13}. 
 However, we identify 
 three groups -- ``NGC3031'', ``ESO507-025'' and ``NGC3311'' with $M_\mathrm{dyn}/\Mbar$ values of $0.23, 1.34$ and $2.91$ respectively -- which contain dark matter much lower than the expected value. While it is essential to remember that NGC3031 and NGC3311 groups were classified as dynamically unrelaxed, the presence of at least one group with minimal dark matter is puzzling. Comprehending this phenomenon becomes difficult when taking into account 
 the merger mechanism commonly used in galaxy formation scenarios. 
More research is needed to reconcile these observations 
 with the prevailing understanding of dark matter's role in galactic dynamics.

Although Equation~\ref{eq:mstar} assumes the galaxies to be early-type galaxies, our results are minimally impacted by the morphology assumed. While \cite{Williams2009} reported the stellar mass-to-light ratio in the $K$-band for late-type galaxies to be $\langle\Upsilon_\mathrm{K}\rangle=1.06\pm0.31$, which is closer to the values observed in this work, \cite{McGaugh2014} found a lower value of 0.6$M_\odot/L_\odot$. To investigate the influence of morphology, we assume a mass-to-light ratio of 0.6$M_\odot/L_\odot$ for all the member galaxies of \citetalias{MK11} to obtain new stellar masses. The vertical residual scatter of \citetalias{MK11} groups about Equation \ref{eq:BFJR} changes from $0.66 \pm 0.32$ dex to $0.41 \pm 0.33$ dex, when the flat velocity dispersion is employed. Thus, morphology is insufficient to account for the observed scatter. It must further be noted that \citetalias{MK11} provide morphology for member galaxies, and only the BCGs of NGC3031, NGC3392, and NGC4631 groups are spirals. As a result, the calculated reduction in scatter here is an overestimation, and our results can be considered unaffected by this assumption.

We emphasise again that the flat dispersion could be lesser than the velocity dispersion in the last bin considered in this study, owing to the declining velocity dispersion profiles. This leads to a greater deviation from the BFJR for these groups towards the left, possibly implying lesser dark matter than the fraction calculated in this work.

The spatially resolved velocity dispersion profiles also offer a test for the MOND paradigm.
The flat velocity dispersions in our subsample-A are systematically smaller than the sample 
 examined in~\cite{Milgrom2019}, as indicated by the crosses in the right panel of Figure~\ref{fig:MVDR}.
Furthermore, since the median baryonic acceleration of our subsample-A, $\gbar\approx5\times10^{-14}$\,m\,s$^{-2}$, 
 is much smaller than $a_0=1.2\times10^{-10}$\,m\,s$^{-2}$ at the last binned radius,
 MOND predicts the BFJR, as shown by Equation~(\ref{eq:BFJR_MOND}) and the green dashed line in Figure~\ref{fig:MVDR}. 
However, the predicted BFJR in MOND is only valid for isolated and spherical systems.
To determine the validity of MOND depends on further testing of its unique prediction 
 regarding the external field effect ~\citep[EFE;][]{Milgrom1983, Haghi2009, Famaey12, Haghi2019, BZ22}.

The non-linear dynamics of MOND in many-body systems, influenced by the EFE, 
 warrant further investigation to test the validity of the MOND paradigm 
 in explaining the observed deviation of the BFJR in galaxy samples.
For instance, when the external field reaches $g_\mathrm{ext}\ge\,a_{0}$, 
 the internal dynamics $g_\mathrm{int}$ approach Newtonian behavior, even in low acceleration regions where $g_\mathrm{int}\ll\,a_{0}$~\citep{Milgrom1983, Milgrom2015, Famaey2018, Muller2019}.
As a result, the deviation of the BFJR in our samples should be tested 
 within the context of the EFE in the MOND paradigm. 
In contrast, it may offer an impressive confirmation when $g_\mathrm{ext}\gg\,a_0$~\citep{Haghi2016, Chae2020, Chae2021}. 
However, this test is beyond the scope of this paper, 
 as it requires detailed information on the background and neighbours of each galaxy group 
 to calculate the total external field $g_\mathrm{ext}$ for analysis.

Surprisingly, MOND offers a specific prediction regarding the insufficient dark matter of the 
 Brightest Group Galaxies (BGGs) in our subsample-A.
Considering the approximate calculation of the baryonic acceleration at the effective radius of the BCG, the Hernquist model~\citep{Hernquist90} indicates that for the NGC4697 group, $\gbar\geq\,a_{0}$.
Little discrepancy is expected between the observed acceleration $\gobs$ and the acceleration due to baryons alone $\gbar$. 
Consequently, BGGs should be explained by Newtonian dynamics with only the baryonic mass, suggesting no dark matter at the centre of galaxy groups.

\section{Summary}\label{sec:conclusion}

The kinematic scaling relation in galaxy groups has been elusive in the literature. It has been believed that the kinematics in galaxy groups/clusters cannot be explained by MOND to be consistent with the BFJR. 
Regardless, \cite{Milgrom2018, Milgrom2019} confirm the dynamics of small galaxy groups to be consistent with MOND by considering the stellar mass only, which implied consistency of galaxy groups with the BFJR.
On the contrary, the MVDR in galaxy clusters was revealed as a parallel BFJR with a larger acceleration scale $\gddag$.
Subsequently, we investigate the spatially resolved kinematic profile in both group and cluster samples to further comprehend the kinematic scaling relation in these systems.

The major results of our work are as follows: 
\begin{enumerate}
    \item The BFJR: By computing the baryonic mass of galaxy groups/clusters and their velocity dispersion profiles, we address an offset from the BFJR expected in MOND for a sample of six clusters and 13 galaxy groups towards the right and left of the relation, respectively.
    Four of the clusters, however, are seen to be consistent with the MVDR. On using the total velocity dispersion, we find that the scatter of \citetalias{MK11} groups about MOND BFJR reduces, whereas six \citetalias{Angus08} clusters are now fully consistent with the MVDR. To assess the validity of MOND, it is necessary to examine the isothermal, spherical symmetry assumptions, and the EFE; 
    \item Two distinct velocity dispersion profiles: Analysis of the velocity dispersion profiles reveals two distinct features: the declining and flat profiles, depending on two parallel BFJRs.
    To validate our results, we test if the systems examined are dynamically relaxed -- all six \citetalias{Angus08} clusters are confirmed to be relaxed, however, five of the 13 \citetalias{MK11} groups are unrelaxed.
\end{enumerate}

Assuming no strong deviations from spherical symmetry, the declining profiles of the Subsample-A indicate deviations from the standard DM profiles such as the isothermal and NFW profiles, where flat velocity dispersion profiles are expected for galaxy groups and clusters. In addition, we discover three galaxy groups with unexpectedly low DM fractions. This presents a paradox: It is difficult to reconcile how galaxy groups, which are formed through the merger of sub-halos abundant in dark matter, could systematically exhibit a deficiency in dark matter.

\section*{Acknowledgements}
We are very grateful to Mordehai Milgrom for his useful suggestions and comments.
We deeply appreciate the anonymous referee's insightful comments, which significantly enhanced this work.
YT is supported by the Taiwan National Science and Technology Council NSTC 110-2112-M-008-015-MY3. 

\section*{Data Availability}
The data used in this work were taken from \cite{Angus08} and \cite{MK11}. Additional data were gathered from the SIMBAD database \citep{simbad} and the NED database.



\bibliographystyle{mnras}
\bibliography{References} 






\bsp	
\label{lastpage}
\end{document}

%% file: Table1.tex
\begin{table*}
  \centering
   \resizebox{2.05\columnwidth}{!}{ \begin{threeparttable}[t]
\begin{tabular}{cccrrcccccc}
  \hline
  Name             & $\mathrm{RA}$\tnote{a} & $\mathrm{DEC}$\tnote{a} & \multicolumn{1}{c}{$N_{\mathrm{gal}}$\tnote{b}} & \multicolumn{1}{c}{$d_A$\tnote{c}}   & $\log(M_{\mathrm{gas}})$\tnote{d} & $\log(M_{\mathrm{star}})$\tnote{d} & $\log(M_{\mathrm{bar}})$\tnote{d} & $\sigma_\mathrm{los, f}$\tnote{e} & $\sigma_\mathrm{los, f}^2/R$\tnote{f} & References     \\
                 & (deg)         & (deg)          & \multicolumn{1}{c}{}                   & \multicolumn{1}{c}{(Mpc)}   & ($M_\odot$)              & ($M_\odot$)               & ($M_\odot$)              & (km/s)            &      $\mathrm{(m/s^2)}$   &       \\
\hline
ACO133           & $01:02:41.8$  & $-21:52:56$    & 45                                     & 239                         & $13.42 \pm 0.08$         & $12.33 \pm 0.03$          & $13.45\pm 0.08 $         & $901\pm240$              & $2.2 \times 10^{-11}$ & (1), (5), (23)       \\
ACO262           & $01:52:46.5$  & $+36:09:07$    & 96                                     & 67.2                        & $13.03 \pm 0.05$         & $12.77 \pm 0.02$          & $13.22\pm 0.03 $         & $631\pm93\enspace$        & $4.2 \times 10^{-12}$ & (1), (4), (5), (24)  \\
ACO383           & $02:48:03.4$  & $-03:31:45$    & 288                                    & 646                         & $13.58 \pm 0.07$         & $12.11 \pm 0.02$          & $13.59\pm 0.07 $         & $510\pm112$               &  $1.0 \times 10^{-12}$  & (1), (5), (25)       \\
ACO1795          & $13:47:22.6$  & $+26:22:52$    & 106                                    & 250                         & $13.80 \pm 0.06$         & $12.52 \pm 0.03$          & $13.82\pm 0.06 $         & $876\pm124$              &   $1.2 \times 10^{-11}$  & (1), (5), (26)       \\
ACO1991          & $14:54:31.5$  & $+18:38:33$    & 63                                     & 236                         & $13.10 \pm 0.09$         & $12.49 \pm 0.02$          & $13.19\pm 0.08 $         & $427\pm103$              &  $4.0 \times 10^{-12}$  & (1), (5), (23), (27)       \\
ACO2589          & $23:23:57.4$  & $+16:46:38$    & 82                                     & 171                         & $13.25 \pm 0.03$         & $12.64 \pm 0.03$          & $13.34\pm 0.02 $         & $786\pm111$              & $1.3 \times 10^{-11}$  & (1), (6), (28)       \\
\hline
NGC3031 Group    & $09:45:33.4$  & $+69:16:12$    & 30                                     & 3.61$^{\dagger}$                     & -                        & $11.22 \pm 0.10$          & $11.22 \pm 0.10$         & $\enspace70\pm17\enspace$ & $4.7 \times10^{-12}$ & (2), (3), (8)  \\
NGC3311 Group    & $10:36:22.2$  & $-27:51:37$    & 139                                    & 52.17                        & -                        & $12.74 \pm 0.02$          & $12.74 \pm 0.02$         & $183\pm38\enspace$   &  $5.3 \times 10^{-13}$    & (2), (3), (9)  \\
NGC3607 Group    & $11:11:47.3$  & $+17:59:46$    & 31                                     & 15.43                        & $\enspace 9.90 \pm 0.05$ & $11.33 \pm 0.07$          & $11.35\pm 0.07 $         & $\enspace54\pm14\enspace$ & $1.2 \times 10^{-13}$   & (2), (3), (10) \\
NGC3992 Group    & $11:54:43.8$  & $+53:10:48$    & 72                                     & 14.61                       & -                        & $11.72 \pm 0.05$          & $11.72 \pm 0.05$         & $123\pm22\enspace$      &  $5.1 \times 10^{-13}$  & (2), (3), (11) \\
NGC4261 Group    & $12:19:37.4$  & $+05:45:12$    & 87                                     & 31.72                       & $10.57 \pm 0.05$         & $12.19 \pm 0.04$          & $12.20\pm 0.04 $         & $179\pm27\enspace$      &  $1.1 \times 10^{-13}$ & (2), (3), (12) \\
NGC4631 Group    & $12:31:39.9$  & $+30:42:37$    & 28                                     & 9.32                       & 8.11                     & $11.24 \pm 0.07$           & $11.24\pm 0.07 $         & $\enspace76\pm29\enspace$ & $6.0 \times 10^{-13}$ & (2), (3), (13) \\
NGC4472 Group    & $12:32:02.1$  & $+11:54:22$    & 355                                    & 15.77                        & $10.67 \pm 0.07$         & $12.67 \pm 0.03$          & $12.67\pm 0.03 $         & $187\pm20\enspace$       & $5.0 \times 10^{-13}$ & (2), (3), (14) \\
NGC4636 Group    & $12:40:51.4$  & $+03:05:05$    & 32                                     & 12.84                       & $11.07 \pm 0.04$         & $11.35 \pm 0.08$          & $11.53\pm 0.05 $         & $\enspace76\pm21\enspace$ & $2.0 \times 10^{-13}$ & (2), (3), (15) \\
NGC4696 Group    & $12:46:26.0$  & $-41:12:12$    & 116                                    & 45.02                       & $11.81 \pm 0.06$         & $12.74 \pm 0.03$          & $12.79\pm 0.03 $         & $181\pm35\enspace$       & $4.7 \times 10^{-13}$ & (2), (3), (16) \\
ESO507-025 Group & $12:51:45.2$  & $-26:31:58$    & 26                                     & 47.24                       & -                        & $12.13 \pm 0.04$           & $12.13 \pm 0.04$         & $\enspace96\pm32\enspace$ &  $3.5 \times 10^{-13}$ & (2), (3), (17) \\
NGC4697 Group    & $12:51:48.1$  & $-07:43:21$    & 37                                     & 19.40                       & $10.54 \pm 0.07$         & $11.94 \pm 0.07$          & $11.96\pm 0.07 $         & $\enspace98\pm27\enspace$ &  $2.9 \times 10^{-13}$ & (2), (3), (18) \\
NGC5044 Group    & $13:14:50.2$  & $-16:25:09$    & 52                                     & 38.45                        & $11.39 \pm 0.05$         & $12.16 \pm 0.04$          & $12.23\pm 0.07 $         & $155\pm33\enspace$        & $6.6 \times 10^{-13}$ & (2), (3), (19) \\
NGC5846 Group    & $15:05:12.2$  & $+01:45:47$    & 74                                     & 26.35                       & $10.82 \pm 0.05$         & $11.95 \pm 0.07$          & $11.98\pm 0.07$           & $117\pm26\enspace$        &  $5.0 \times 10^{-13}$ & (2), (3), (20)\\
  \hline
  \end{tabular}
  \caption{The properties of the sample of groups and clusters.}
  \vspace{-8pt}
  \begin{tablenotes}
  \textit{Notes$-$} The first six rows (above the horizontal line) are clusters from \protect\citetalias{Angus08} and the remaining groups below the horizontal line are from \protect\citetalias{MK11} (named after their brightest galaxy). The column description are as follows:\\
    \item[a] RA and DEC are the Right Ascension and Declinations of the BCG for \protect\citetalias{Angus08} clusters, whereas they stand for the centroid of the groups for \protect\citetalias{MK11} groups.\\
    \item[b] Number of member galaxies of the respective group/cluster (richness).\\
    \item[c] Angular diameter distance estimated from the redshift of group/cluster.\\
    \item[d] $M_\mathrm{gas}$ stands for the gas mass, and $M_\mathrm{star}$ stands for the stellar mass calculated from Equation (\protect\ref{eq:mstar}), respectively of the group/cluster. Combining the stellar and gas masses, we estimate the baryonic mass $M_\mathrm{bar}$.\\
    \item[e] The velocity dispersion in the last bin, estimated using biweight estimator, which is considered the flat velocity dispersion in this work.\\
    \item[f] An estimate of the acceleration of these systems in the regions where $\sigma_\mathrm{los,f}$ is determined. $R$ is the projected radius of the last bin in which $\sigma_\mathrm{los,f}$ is computed. This estimate shows that these systems are in deep-MOND limit, with accelerations smaller than $a_0 = 1.2 \times 10^{-10}\,\mathrm{m/s^2}$. In addition, the galaxy groups have accelerations smaller than the clusters by more than one order of magnitude, because of their extended size.\\
    \item[$\dagger$] The distance for this group has been taken from NED database because of its proximity to the Milky Way.\\
    \textit{References} $-$ (1) \protect\cite{Angus08}; (2) \protect\cite{MK11}; (3) \protect\cite{Paturel2003}; (4) \protect\cite{Gastaldello07a}; (5) \protect\cite{Vikhlinin2006}; (6) \protect\cite{Zappacosta2006}; (7) \protect\cite{2mass};
(8) \protect\cite{8t}; (9) \protect\cite{9t}; (10) \protect\cite{10t}; (11) \protect\cite{11t}; (12) \protect\cite{12t}; (13) \protect\cite{13t}; (14) \protect\cite{14t}; (15) \protect\cite{15t}; (16) \protect\cite{16t}; (17) \protect\cite{17t}; (18) \protect\cite{18t}; (19) \protect\cite{19t}; (20) \protect\cite{20t}; (21) \protect\cite{Babyk18}; (22) \protect\cite{yamasaki2009}; (23) \protect\cite{Smith2004}; (24) \protect\cite{Sakai2012}; (25) \protect\cite{Geller2014}; (26) \protect\cite{Cava2009}; (27) \protect\cite{Marziani2017}; (28) \protect\cite{Oh2018}
  \end{tablenotes}
  \label{tab:table1}
 \end{threeparttable}}
\end{table*}

%% file: BFJR_GG.bbl
\begin{thebibliography}{}
\makeatletter
\relax
\def\mn@urlcharsother{\let\do\@makeother \do\$\do\&\do\#\do\^\do\_\do\%\do\~}
\def\mn@doi{\begingroup\mn@urlcharsother \@ifnextchar [ {\mn@doi@}
  {\mn@doi@[]}}
\def\mn@doi@[#1]#2{\def\@tempa{#1}\ifx\@tempa\@empty \href
  {http://dx.doi.org/#2} {doi:#2}\else \href {http://dx.doi.org/#2} {#1}\fi
  \endgroup}
\def\mn@eprint#1#2{\mn@eprint@#1:#2::\@nil}
\def\mn@eprint@arXiv#1{\href {http://arxiv.org/abs/#1} {{\tt arXiv:#1}}}
\def\mn@eprint@dblp#1{\href {http://dblp.uni-trier.de/rec/bibtex/#1.xml}
  {dblp:#1}}
\def\mn@eprint@#1:#2:#3:#4\@nil{\def\@tempa {#1}\def\@tempb {#2}\def\@tempc
  {#3}\ifx \@tempc \@empty \let \@tempc \@tempb \let \@tempb \@tempa \fi \ifx
  \@tempb \@empty \def\@tempb {arXiv}\fi \@ifundefined
  {mn@eprint@\@tempb}{\@tempb:\@tempc}{\expandafter \expandafter \csname
  mn@eprint@\@tempb\endcsname \expandafter{\@tempc}}}

\bibitem[\protect\citeauthoryear{{Andreon}}{{Andreon}}{2010}]{Andreon2010}
{Andreon} S.,  2010, \mn@doi [\mnras] {10.1111/j.1365-2966.2010.16856.x}, \href
  {https://ui.adsabs.harvard.edu/abs/2010MNRAS.407..263A} {407, 263}

\bibitem[\protect\citeauthoryear{{Angus}, {Famaey}  \& {Buote}}{{Angus}
  et~al.}{2008}]{Angus08}
{Angus} G.~W.,  {Famaey} B.,   {Buote} D.~A.,  2008, \mn@doi [\mnras]
  {10.1111/j.1365-2966.2008.13353.x}, \href
  {https://ui.adsabs.harvard.edu/abs/2008MNRAS.387.1470A} {387, 1470}

\bibitem[\protect\citeauthoryear{{Babyk}, {McNamara}, {Nulsen}, {Hogan},
  {Vantyghem}, {Russell}, {Pulido}  \& {Edge}}{{Babyk} et~al.}{2018}]{Babyk18}
{Babyk} I.~V.,  {McNamara} B.~R.,  {Nulsen} P.~E.~J.,  {Hogan} M.~T.,
  {Vantyghem} A.~N.,  {Russell} H.~R.,  {Pulido} F.~A.,   {Edge} A.~C.,  2018,
  \mn@doi [\apj] {10.3847/1538-4357/aab3c9}, \href
  {https://ui.adsabs.harvard.edu/abs/2018ApJ...857...32B} {857, 32}

\bibitem[\protect\citeauthoryear{{Bahar} et~al.,}{{Bahar}
  et~al.}{2022}]{Bahar2022}
{Bahar} Y.~E.,  et~al., 2022, \mn@doi [\aap] {10.1051/0004-6361/202142462},
  \href {https://ui.adsabs.harvard.edu/abs/2022A&A...661A...7B} {661, A7}

\bibitem[\protect\citeauthoryear{{Banik} \& {Zhao}}{{Banik} \&
  {Zhao}}{2022}]{BZ22}
{Banik} I.,  {Zhao} H.,  2022, \mn@doi [Symmetry] {10.3390/sym14071331}, \href
  {https://ui.adsabs.harvard.edu/abs/2022Symm...14.1331B} {14, 1331}

\bibitem[\protect\citeauthoryear{{Beers}, {Flynn}  \& {Gebhardt}}{{Beers}
  et~al.}{1990}]{Beers1990}
{Beers} T.~C.,  {Flynn} K.,   {Gebhardt} K.,  1990, \mn@doi [\aj]
  {10.1086/115487}, \href
  {https://ui.adsabs.harvard.edu/abs/1990AJ....100...32B} {100, 32}

\bibitem[\protect\citeauthoryear{{Bell}, {McIntosh}, {Katz}  \&
  {Weinberg}}{{Bell} et~al.}{2003}]{Bell2003}
{Bell} E.~F.,  {McIntosh} D.~H.,  {Katz} N.,   {Weinberg} M.~D.,  2003, \mn@doi
  [\apjs] {10.1086/378847}, \href
  {https://ui.adsabs.harvard.edu/abs/2003ApJS..149..289B} {149, 289}

\bibitem[\protect\citeauthoryear{{Berezhiani}, {Famaey}  \&
  {Khoury}}{{Berezhiani} et~al.}{2018}]{Berezhiani2018}
{Berezhiani} L.,  {Famaey} B.,   {Khoury} J.,  2018, \mn@doi [\jcap]
  {10.1088/1475-7516/2018/09/021}, \href
  {https://ui.adsabs.harvard.edu/abs/2018JCAP...09..021B} {2018, 021}

\bibitem[\protect\citeauthoryear{{Bergond}, {Zepf}, {Romanowsky}, {Sharples}
  \& {Rhode}}{{Bergond} et~al.}{2006}]{Bergond2006}
{Bergond} G.,  {Zepf} S.~E.,  {Romanowsky} A.~J.,  {Sharples} R.~M.,   {Rhode}
  K.~L.,  2006, \mn@doi [\aap] {10.1051/0004-6361:20053697}, \href
  {https://ui.adsabs.harvard.edu/abs/2006A&A...448..155B} {448, 155}

\bibitem[\protect\citeauthoryear{{Bernardi}, {Alonso}, {da Costa}, {Willmer},
  {Wegner}, {Pellegrini}, {Rit{\'e}}  \& {Maia}}{{Bernardi} et~al.}{2002}]{17t}
{Bernardi} M.,  {Alonso} M.~V.,  {da Costa} L.~N.,  {Willmer} C.~N.~A.,
  {Wegner} G.,  {Pellegrini} P.~S.,  {Rit{\'e}} C.,   {Maia} M.~A.~G.,  2002,
  \mn@doi [\aj] {10.1086/340463}, \href
  {https://ui.adsabs.harvard.edu/abs/2002AJ....123.2990B} {123, 2990}

\bibitem[\protect\citeauthoryear{{Binney} \& {Tremaine}}{{Binney} \&
  {Tremaine}}{2008}]{BT08}
{Binney} J.,  {Tremaine} S.,  2008, {Galactic Dynamics: Second Edition}.
Princeton University Press

\bibitem[\protect\citeauthoryear{{Capelo}, {Coppi}  \& {Natarajan}}{{Capelo}
  et~al.}{2012}]{Capelo2012}
{Capelo} P.~R.,  {Coppi} P.~S.,   {Natarajan} P.,  2012, \mn@doi [\mnras]
  {10.1111/j.1365-2966.2012.20648.x}, \href
  {https://ui.adsabs.harvard.edu/abs/2012MNRAS.422..686C} {422, 686}

\bibitem[\protect\citeauthoryear{{Cappellari}}{{Cappellari}}{2013}]{Capellari2013}
{Cappellari} M.,  2013, \mn@doi [\apjl] {10.1088/2041-8205/778/1/L2}, \href
  {https://ui.adsabs.harvard.edu/abs/2013ApJ...778L...2C} {778, L2}

\bibitem[\protect\citeauthoryear{{Cava} et~al.,}{{Cava}
  et~al.}{2009}]{Cava2009}
{Cava} A.,  et~al., 2009, \mn@doi [\aap] {10.1051/0004-6361:200810997}, \href
  {https://ui.adsabs.harvard.edu/abs/2009A&A...495..707C} {495, 707}

\bibitem[\protect\citeauthoryear{{Chae}, {Lelli}, {Desmond}, {McGaugh}, {Li}
  \& {Schombert}}{{Chae} et~al.}{2020}]{Chae2020}
{Chae} K.-H.,  {Lelli} F.,  {Desmond} H.,  {McGaugh} S.~S.,  {Li} P.,
  {Schombert} J.~M.,  2020, \mn@doi [\apj] {10.3847/1538-4357/abbb96}, \href
  {https://ui.adsabs.harvard.edu/abs/2020ApJ...904...51C} {904, 51}

\bibitem[\protect\citeauthoryear{{Chae}, {Desmond}, {Lelli}, {McGaugh}  \&
  {Schombert}}{{Chae} et~al.}{2021}]{Chae2021}
{Chae} K.-H.,  {Desmond} H.,  {Lelli} F.,  {McGaugh} S.~S.,   {Schombert}
  J.~M.,  2021, \mn@doi [\apj] {10.3847/1538-4357/ac1bba}, \href
  {https://ui.adsabs.harvard.edu/abs/2021ApJ...921..104C} {921, 104}

\bibitem[\protect\citeauthoryear{{Chan} \& {Del Popolo}}{{Chan} \& {Del
  Popolo}}{2020}]{Chan2020}
{Chan} M.~H.,  {Del Popolo} A.,  2020, \mn@doi [\mnras]
  {10.1093/mnras/staa225}, \href
  {https://ui.adsabs.harvard.edu/abs/2020MNRAS.492.5865C} {492, 5865}

\bibitem[\protect\citeauthoryear{{Choi}, {Park}  \& {Vogeley}}{{Choi}
  et~al.}{2007}]{Choi2007}
{Choi} Y.-Y.,  {Park} C.,   {Vogeley} M.~S.,  2007, \mn@doi [\apj]
  {10.1086/511060}, \href
  {https://ui.adsabs.harvard.edu/abs/2007ApJ...658..884C} {658, 884}

\bibitem[\protect\citeauthoryear{{Costa}, {Ribeiro}  \& {de Carvalho}}{{Costa}
  et~al.}{2018}]{Costa18}
{Costa} A.~P.,  {Ribeiro} A.~L.~B.,   {de Carvalho} R.~R.,  2018, \mn@doi
  [\mnras] {10.1093/mnrasl/slx156}, \href
  {https://ui.adsabs.harvard.edu/abs/2018MNRAS.473L..31C} {473, L31}

\bibitem[\protect\citeauthoryear{{Courteau}, {Dutton}, {van den Bosch},
  {MacArthur}, {Dekel}, {McIntosh}  \& {Dale}}{{Courteau}
  et~al.}{2007}]{Courteau2007}
{Courteau} S.,  {Dutton} A.~A.,  {van den Bosch} F.~C.,  {MacArthur} L.~A.,
  {Dekel} A.,  {McIntosh} D.~H.,   {Dale} D.~A.,  2007, \mn@doi [\apj]
  {10.1086/522193}, \href
  {https://ui.adsabs.harvard.edu/abs/2007ApJ...671..203C} {671, 203}

\bibitem[\protect\citeauthoryear{{Dai}, {Bregman}, {Kochanek}  \&
  {Rasia}}{{Dai} et~al.}{2010}]{Dai10}
{Dai} X.,  {Bregman} J.~N.,  {Kochanek} C.~S.,   {Rasia} E.,  2010, \mn@doi
  [\apj] {10.1088/0004-637X/719/1/119}, \href
  {https://ui.adsabs.harvard.edu/abs/2010ApJ...719..119D} {719, 119}

\bibitem[\protect\citeauthoryear{{Desmond}}{{Desmond}}{2017}]{Desmond2017b}
{Desmond} H.,  2017, \mn@doi [\mnras] {10.1093/mnrasl/slx134}, \href
  {https://ui.adsabs.harvard.edu/abs/2017MNRAS.472L..35D} {472, L35}

\bibitem[\protect\citeauthoryear{{Desmond} \& {Wechsler}}{{Desmond} \&
  {Wechsler}}{2015}]{Desmond2015}
{Desmond} H.,  {Wechsler} R.~H.,  2015, \mn@doi [\mnras]
  {10.1093/mnras/stv1978}, \href
  {https://ui.adsabs.harvard.edu/abs/2015MNRAS.454..322D} {454, 322}

\bibitem[\protect\citeauthoryear{{Desmond} \& {Wechsler}}{{Desmond} \&
  {Wechsler}}{2017}]{Desmond2017a}
{Desmond} H.,  {Wechsler} R.~H.,  2017, \mn@doi [\mnras]
  {10.1093/mnras/stw2804}, \href
  {https://ui.adsabs.harvard.edu/abs/2017MNRAS.465..820D} {465, 820}

\bibitem[\protect\citeauthoryear{{Desroches}, {Quataert}, {Ma}  \&
  {West}}{{Desroches} et~al.}{2007}]{Desroches2007}
{Desroches} L.-B.,  {Quataert} E.,  {Ma} C.-P.,   {West} A.~A.,  2007, \mn@doi
  [\mnras] {10.1111/j.1365-2966.2007.11612.x}, \href
  {https://ui.adsabs.harvard.edu/abs/2007MNRAS.377..402D} {377, 402}

\bibitem[\protect\citeauthoryear{{Djorgovski} \& {Davis}}{{Djorgovski} \&
  {Davis}}{1987}]{Djorgovski1987}
{Djorgovski} S.,  {Davis} M.,  1987, \mn@doi [\apj] {10.1086/164948}, \href
  {https://ui.adsabs.harvard.edu/abs/1987ApJ...313...59D} {313, 59}

\bibitem[\protect\citeauthoryear{{Dressler}, {Lynden-Bell}, {Burstein},
  {Davies}, {Faber}, {Terlevich}  \& {Wegner}}{{Dressler}
  et~al.}{1987}]{Dressler1987}
{Dressler} A.,  {Lynden-Bell} D.,  {Burstein} D.,  {Davies} R.~L.,  {Faber}
  S.~M.,  {Terlevich} R.,   {Wegner} G.,  1987, \mn@doi [\apj]
  {10.1086/164947}, \href
  {https://ui.adsabs.harvard.edu/abs/1987ApJ...313...42D} {313, 42}

\bibitem[\protect\citeauthoryear{{Durazo}, {Hernandez}, {Cervantes Sodi}  \&
  {S{\'a}nchez}}{{Durazo} et~al.}{2017}]{Durazo2017}
{Durazo} R.,  {Hernandez} X.,  {Cervantes Sodi} B.,   {S{\'a}nchez} S.~F.,
  2017, \mn@doi [\apj] {10.3847/1538-4357/aa619a}, \href
  {https://ui.adsabs.harvard.edu/abs/2017ApJ...837..179D} {837, 179}

\bibitem[\protect\citeauthoryear{{Dutton} \& {van den Bosch}}{{Dutton} \& {van
  den Bosch}}{2009}]{Dutton2009}
{Dutton} A.~A.,  {van den Bosch} F.~C.,  2009, \mn@doi [\mnras]
  {10.1111/j.1365-2966.2009.14742.x}, \href
  {https://ui.adsabs.harvard.edu/abs/2009MNRAS.396..141D} {396, 141}

\bibitem[\protect\citeauthoryear{{Eckert}, {Ettori}, {Pointecouteau},
  {Molendi}, {Paltani}  \& {Tchernin}}{{Eckert} et~al.}{2017}]{Eckert2017}
{Eckert} D.,  {Ettori} S.,  {Pointecouteau} E.,  {Molendi} S.,  {Paltani} S.,
  {Tchernin} C.,  2017, \mn@doi [Astronomische Nachrichten]
  {10.1002/asna.201713345}, \href
  {https://ui.adsabs.harvard.edu/abs/2017AN....338..293E} {338, 293}

\bibitem[\protect\citeauthoryear{{Eckert}, {Ettori}, {Pointecouteau}, {van der
  Burg}  \& {Loubser}}{{Eckert} et~al.}{2022}]{Ettori22}
{Eckert} D.,  {Ettori} S.,  {Pointecouteau} E.,  {van der Burg} R.~F.~J.,
  {Loubser} S.~I.,  2022, \mn@doi [\aap] {10.1051/0004-6361/202142507}, \href
  {https://ui.adsabs.harvard.edu/abs/2022A&A...662A.123E} {662, A123}

\bibitem[\protect\citeauthoryear{{Faber} \& {Jackson}}{{Faber} \&
  {Jackson}}{1976}]{Faber1976}
{Faber} S.~M.,  {Jackson} R.~E.,  1976, \mn@doi [\apj] {10.1086/154215}, \href
  {https://ui.adsabs.harvard.edu/abs/1976ApJ...204..668F} {204, 668}

\bibitem[\protect\citeauthoryear{{Fabian}}{{Fabian}}{2012}]{Fabian2012}
{Fabian} A.~C.,  2012, \mn@doi [\araa] {10.1146/annurev-astro-081811-125521},
  \href {https://ui.adsabs.harvard.edu/abs/2012ARA&A..50..455F} {50, 455}

\bibitem[\protect\citeauthoryear{{Famaey} \& {McGaugh}}{{Famaey} \&
  {McGaugh}}{2012}]{Famaey12}
{Famaey} B.,  {McGaugh} S.~S.,  2012, \mn@doi [Living Reviews in Relativity]
  {10.12942/lrr-2012-10}, \href
  {https://ui.adsabs.harvard.edu/abs/2012LRR....15...10F} {15, 10}

\bibitem[\protect\citeauthoryear{{Famaey}, {McGaugh}  \& {Milgrom}}{{Famaey}
  et~al.}{2018}]{Famaey2018}
{Famaey} B.,  {McGaugh} S.,   {Milgrom} M.,  2018, \mn@doi [\mnras]
  {10.1093/mnras/sty1884}, \href
  {https://ui.adsabs.harvard.edu/abs/2018MNRAS.480..473F} {480, 473}

\bibitem[\protect\citeauthoryear{{Ferrarese} et~al.,}{{Ferrarese}
  et~al.}{2000}]{8t}
{Ferrarese} L.,  et~al., 2000, \mn@doi [\apjs] {10.1086/313391}, \href
  {https://ui.adsabs.harvard.edu/abs/2000ApJS..128..431F} {128, 431}

\bibitem[\protect\citeauthoryear{{Gastaldello}, {Buote}, {Humphrey},
  {Zappacosta}, {Bullock}, {Brighenti}  \& {Mathews}}{{Gastaldello}
  et~al.}{2007}]{Gastaldello07a}
{Gastaldello} F.,  {Buote} D.~A.,  {Humphrey} P.~J.,  {Zappacosta} L.,
  {Bullock} J.~S.,  {Brighenti} F.,   {Mathews} W.~G.,  2007, \mn@doi [\apj]
  {10.1086/521519}, \href
  {https://ui.adsabs.harvard.edu/abs/2007ApJ...669..158G} {669, 158}

\bibitem[\protect\citeauthoryear{{Geller}, {Hwang}, {Diaferio}, {Kurtz}, {Coe}
  \& {Rines}}{{Geller} et~al.}{2014}]{Geller2014}
{Geller} M.~J.,  {Hwang} H.~S.,  {Diaferio} A.,  {Kurtz} M.~J.,  {Coe} D.,
  {Rines} K.~J.,  2014, \mn@doi [\apj] {10.1088/0004-637X/783/1/52}, \href
  {https://ui.adsabs.harvard.edu/abs/2014ApJ...783...52G} {783, 52}

\bibitem[\protect\citeauthoryear{{Ghari}, {Haghi}  \& {Zonoozi}}{{Ghari}
  et~al.}{2019}]{Ghari2019}
{Ghari} A.,  {Haghi} H.,   {Zonoozi} A.~H.,  2019, \mn@doi [\mnras]
  {10.1093/mnras/stz1272}, \href
  {https://ui.adsabs.harvard.edu/abs/2019MNRAS.487.2148G} {487, 2148}

\bibitem[\protect\citeauthoryear{{Giodini} et~al.,}{{Giodini}
  et~al.}{2009}]{Giodini09}
{Giodini} S.,  et~al., 2009, \mn@doi [\apj] {10.1088/0004-637X/703/1/982},
  \href {https://ui.adsabs.harvard.edu/abs/2009ApJ...703..982G} {703, 982}

\bibitem[\protect\citeauthoryear{{Gonzalez}, {Zaritsky}  \&
  {Zabludoff}}{{Gonzalez} et~al.}{2007}]{Gonzalez2007}
{Gonzalez} A.~H.,  {Zaritsky} D.,   {Zabludoff} A.~I.,  2007, \mn@doi [\apj]
  {10.1086/519729}, \href
  {https://ui.adsabs.harvard.edu/abs/2007ApJ...666..147G} {666, 147}

\bibitem[\protect\citeauthoryear{{Gonzalez}, {Sivanandam}, {Zabludoff}  \&
  {Zaritsky}}{{Gonzalez} et~al.}{2013}]{Gonzalez13}
{Gonzalez} A.~H.,  {Sivanandam} S.,  {Zabludoff} A.~I.,   {Zaritsky} D.,  2013,
  \mn@doi [\apj] {10.1088/0004-637X/778/1/14}, \href
  {https://ui.adsabs.harvard.edu/abs/2013ApJ...778...14G} {778, 14}

\bibitem[\protect\citeauthoryear{{Groener}, {Goldberg}  \& {Sereno}}{{Groener}
  et~al.}{2016}]{20t}
{Groener} A.~M.,  {Goldberg} D.~M.,   {Sereno} M.,  2016, \mn@doi [\mnras]
  {10.1093/mnras/stv2341}, \href
  {https://ui.adsabs.harvard.edu/abs/2016MNRAS.455..892G} {455, 892}

\bibitem[\protect\citeauthoryear{{Haghi}, {Baumgardt}, {Kroupa}, {Grebel},
  {Hilker}  \& {Jordi}}{{Haghi} et~al.}{2009}]{Haghi2009}
{Haghi} H.,  {Baumgardt} H.,  {Kroupa} P.,  {Grebel} E.~K.,  {Hilker} M.,
  {Jordi} K.,  2009, \mn@doi [\mnras] {10.1111/j.1365-2966.2009.14656.x}, \href
  {https://ui.adsabs.harvard.edu/abs/2009MNRAS.395.1549H} {395, 1549}

\bibitem[\protect\citeauthoryear{{Haghi}, {Bazkiaei}, {Zonoozi}  \&
  {Kroupa}}{{Haghi} et~al.}{2016}]{Haghi2016}
{Haghi} H.,  {Bazkiaei} A.~E.,  {Zonoozi} A.~H.,   {Kroupa} P.,  2016, \mn@doi
  [\mnras] {10.1093/mnras/stw573}, \href
  {https://ui.adsabs.harvard.edu/abs/2016MNRAS.458.4172H} {458, 4172}

\bibitem[\protect\citeauthoryear{{Haghi} et~al.,}{{Haghi}
  et~al.}{2019}]{Haghi2019}
{Haghi} H.,  et~al., 2019, \mn@doi [\mnras] {10.1093/mnras/stz1465}, \href
  {https://ui.adsabs.harvard.edu/abs/2019MNRAS.487.2441H} {487, 2441}

\bibitem[\protect\citeauthoryear{{Helsdon} \& {Ponman}}{{Helsdon} \&
  {Ponman}}{2000}]{15t}
{Helsdon} S.~F.,  {Ponman} T.~J.,  2000, \mn@doi [\mnras]
  {10.1046/j.1365-8711.2000.03396.x}, \href
  {https://ui.adsabs.harvard.edu/abs/2000MNRAS.315..356H} {315, 356}

\bibitem[\protect\citeauthoryear{{Henriksen}}{{Henriksen}}{2011}]{19t}
{Henriksen} M.~J.,  2011, \mn@doi [\apj] {10.1088/0004-637X/726/1/9}, \href
  {https://ui.adsabs.harvard.edu/abs/2011ApJ...726....9H} {726, 9}

\bibitem[\protect\citeauthoryear{{Hernandez} \& {Jim{\'e}nez}}{{Hernandez} \&
  {Jim{\'e}nez}}{2012}]{Hernandez2012}
{Hernandez} X.,  {Jim{\'e}nez} M.~A.,  2012, \mn@doi [\apj]
  {10.1088/0004-637X/750/1/9}, \href
  {https://ui.adsabs.harvard.edu/abs/2012ApJ...750....9H} {750, 9}

\bibitem[\protect\citeauthoryear{{Hernquist}}{{Hernquist}}{1990}]{Hernquist90}
{Hernquist} L.,  1990, \mn@doi [\apj] {10.1086/168845}, \href
  {https://ui.adsabs.harvard.edu/abs/1990ApJ...356..359H} {356, 359}

\bibitem[\protect\citeauthoryear{{Hodson} \& {Zhao}}{{Hodson} \&
  {Zhao}}{2017}]{HZ17}
{Hodson} A.~O.,  {Zhao} H.,  2017, \mn@doi [\aap]
  {10.1051/0004-6361/201629358}, \href
  {https://ui.adsabs.harvard.edu/abs/2017A&A...598A.127H} {598, A127}

\bibitem[\protect\citeauthoryear{{Hou}, {Parker}, {Harris}  \& {Wilman}}{{Hou}
  et~al.}{2009}]{Annie2009}
{Hou} A.,  {Parker} L.~C.,  {Harris} W.~E.,   {Wilman} D.~J.,  2009, \mn@doi
  [\apj] {10.1088/0004-637X/702/2/1199}, \href
  {https://ui.adsabs.harvard.edu/abs/2009ApJ...702.1199H} {702, 1199}

\bibitem[\protect\citeauthoryear{{Karachentsev} \& {Makarov}}{{Karachentsev} \&
  {Makarov}}{1996}]{Karachentsev1996}
{Karachentsev} I.~D.,  {Makarov} D.~A.,  1996, \mn@doi [\aj] {10.1086/117825},
  \href {https://ui.adsabs.harvard.edu/abs/1996AJ....111..794K} {111, 794}

\bibitem[\protect\citeauthoryear{{Kauffmann} \& {Charlot}}{{Kauffmann} \&
  {Charlot}}{1998}]{Kauffmann98}
{Kauffmann} G.,  {Charlot} S.,  1998, \mn@doi [\mnras]
  {10.1046/j.1365-8711.1998.01708.x}, \href
  {https://ui.adsabs.harvard.edu/abs/1998MNRAS.297L..23K} {297, L23}

\bibitem[\protect\citeauthoryear{{Lelli}, {McGaugh}  \& {Schombert}}{{Lelli}
  et~al.}{2016}]{Lelli2016}
{Lelli} F.,  {McGaugh} S.~S.,   {Schombert} J.~M.,  2016, \mn@doi [\apjl]
  {10.3847/2041-8205/816/1/L14}, \href
  {https://ui.adsabs.harvard.edu/abs/2016ApJ...816L..14L} {816, L14}

\bibitem[\protect\citeauthoryear{{Lelli}, {McGaugh}, {Schombert}  \&
  {Pawlowski}}{{Lelli} et~al.}{2017}]{Lelli2017}
{Lelli} F.,  {McGaugh} S.~S.,  {Schombert} J.~M.,   {Pawlowski} M.~S.,  2017,
  \mn@doi [\apj] {10.3847/1538-4357/836/2/152}, \href
  {https://ui.adsabs.harvard.edu/abs/2017ApJ...836..152L} {836, 152}

\bibitem[\protect\citeauthoryear{{Lelli}, {McGaugh}, {Schombert}, {Desmond}  \&
  {Katz}}{{Lelli} et~al.}{2019}]{Lelli2019}
{Lelli} F.,  {McGaugh} S.~S.,  {Schombert} J.~M.,  {Desmond} H.,   {Katz} H.,
  2019, \mn@doi [\mnras] {10.1093/mnras/stz205}, \href
  {https://ui.adsabs.harvard.edu/abs/2019MNRAS.484.3267L} {484, 3267}

\bibitem[\protect\citeauthoryear{{Li}, {Jing}, {Mao}, {Han}, {Peng}, {Yang},
  {Mo}  \& {van den Bosch}}{{Li} et~al.}{2012}]{Li12}
{Li} C.,  {Jing} Y.~P.,  {Mao} S.,  {Han} J.,  {Peng} Q.,  {Yang} X.,  {Mo}
  H.~J.,   {van den Bosch} F.,  2012, \mn@doi [\apj]
  {10.1088/0004-637X/758/1/50}, \href
  {https://ui.adsabs.harvard.edu/abs/2012ApJ...758...50L} {758, 50}

\bibitem[\protect\citeauthoryear{{Li}, {Lelli}, {McGaugh}  \& {Schombert}}{{Li}
  et~al.}{2018}]{Li2018}
{Li} P.,  {Lelli} F.,  {McGaugh} S.,   {Schombert} J.,  2018, \mn@doi [\aap]
  {10.1051/0004-6361/201732547}, \href
  {https://ui.adsabs.harvard.edu/abs/2018A&A...615A...3L} {615, A3}

\bibitem[\protect\citeauthoryear{{Li}, {McGaugh}, {Lelli}, {Tian}, {Schombert}
  \& {Ko}}{{Li} et~al.}{2022}]{Li2022}
{Li} P.,  {McGaugh} S.~S.,  {Lelli} F.,  {Tian} Y.,  {Schombert} J.~M.,   {Ko}
  C.-M.,  2022, \mn@doi [\apj] {10.3847/1538-4357/ac52aa}, \href
  {https://ui.adsabs.harvard.edu/abs/2022ApJ...927..198L} {927, 198}

\bibitem[\protect\citeauthoryear{{Li} et~al.,}{{Li} et~al.}{2023}]{Li2023}
{Li} P.,  et~al., 2023, \mn@doi [arXiv e-prints] {10.48550/arXiv.2303.10175},
  \href {https://ui.adsabs.harvard.edu/abs/2023arXiv230310175L} {p.
  arXiv:2303.10175}

\bibitem[\protect\citeauthoryear{{Lin} \& {Mohr}}{{Lin} \&
  {Mohr}}{2004}]{Lin04}
{Lin} Y.-T.,  {Mohr} J.~J.,  2004, \mn@doi [\apj] {10.1086/425412}, \href
  {https://ui.adsabs.harvard.edu/abs/2004ApJ...617..879L} {617, 879}

\bibitem[\protect\citeauthoryear{{Liu} et~al.,}{{Liu} et~al.}{2023}]{Liu2023}
{Liu} A.,  et~al., 2023, \mn@doi [\aap] {10.1051/0004-6361/202245118}, \href
  {https://ui.adsabs.harvard.edu/abs/2023A&A...670A..96L} {670, A96}

\bibitem[\protect\citeauthoryear{{{\L}okas} \& {Mamon}}{{{\L}okas} \&
  {Mamon}}{2001}]{Lokas2001}
{{\L}okas} E.~L.,  {Mamon} G.~A.,  2001, \mn@doi [\mnras]
  {10.1046/j.1365-8711.2001.04007.x}, \href
  {https://ui.adsabs.harvard.edu/abs/2001MNRAS.321..155L} {321, 155}

\bibitem[\protect\citeauthoryear{{Lovisari}, {Reiprich}  \&
  {Schellenberger}}{{Lovisari} et~al.}{2015}]{Lovisari2015}
{Lovisari} L.,  {Reiprich} T.~H.,   {Schellenberger} G.,  2015, \mn@doi [\aap]
  {10.1051/0004-6361/201423954}, \href
  {https://ui.adsabs.harvard.edu/abs/2015A&A...573A.118L} {573, A118}

\bibitem[\protect\citeauthoryear{{Lovisari}, {Ettori}, {Gaspari}  \&
  {Giles}}{{Lovisari} et~al.}{2021}]{Lovisari2021}
{Lovisari} L.,  {Ettori} S.,  {Gaspari} M.,   {Giles} P.~A.,  2021, \mn@doi
  [Universe] {10.3390/universe7050139}, \href
  {https://ui.adsabs.harvard.edu/abs/2021Univ....7..139L} {7, 139}

\bibitem[\protect\citeauthoryear{{Makarov} \& {Karachentsev}}{{Makarov} \&
  {Karachentsev}}{2011}]{MK11}
{Makarov} D.,  {Karachentsev} I.,  2011, \mn@doi [\mnras]
  {10.1111/j.1365-2966.2010.18071.x}, \href
  {https://ui.adsabs.harvard.edu/abs/2011MNRAS.412.2498M} {412, 2498}

\bibitem[\protect\citeauthoryear{{Marziani} et~al.,}{{Marziani}
  et~al.}{2017}]{Marziani2017}
{Marziani} P.,  et~al., 2017, \mn@doi [\aap] {10.1051/0004-6361/201628941},
  \href {https://ui.adsabs.harvard.edu/abs/2017A&A...599A..83M} {599, A83}

\bibitem[\protect\citeauthoryear{{McGaugh}}{{McGaugh}}{2011}]{McGaugh2011}
{McGaugh} S.~S.,  2011, \mn@doi [\prl] {10.1103/PhysRevLett.106.121303}, \href
  {https://ui.adsabs.harvard.edu/abs/2011PhRvL.106l1303M} {106, 121303}

\bibitem[\protect\citeauthoryear{{McGaugh}}{{McGaugh}}{2012}]{McGaugh2012}
{McGaugh} S.~S.,  2012, \mn@doi [\aj] {10.1088/0004-6256/143/2/40}, \href
  {https://ui.adsabs.harvard.edu/abs/2012AJ....143...40M} {143, 40}

\bibitem[\protect\citeauthoryear{{McGaugh}}{{McGaugh}}{2020}]{McGaugh20}
{McGaugh} S.~S.,  2020, \mn@doi [Galaxies] {10.3390/galaxies8020035}, \href
  {https://ui.adsabs.harvard.edu/abs/2020Galax...8...35M} {8, 35}

\bibitem[\protect\citeauthoryear{{McGaugh} \& {Schombert}}{{McGaugh} \&
  {Schombert}}{2014}]{McGaugh2014}
{McGaugh} S.~S.,  {Schombert} J.~M.,  2014, \mn@doi [\aj]
  {10.1088/0004-6256/148/5/77}, \href
  {https://ui.adsabs.harvard.edu/abs/2014AJ....148...77M} {148, 77}

\bibitem[\protect\citeauthoryear{{McGaugh}, {Schombert}, {Bothun}  \& {de
  Blok}}{{McGaugh} et~al.}{2000}]{McGaugh2000}
{McGaugh} S.~S.,  {Schombert} J.~M.,  {Bothun} G.~D.,   {de Blok} W.~J.~G.,
  2000, \mn@doi [\apjl] {10.1086/312628}, \href
  {https://ui.adsabs.harvard.edu/abs/2000ApJ...533L..99M} {533, L99}

\bibitem[\protect\citeauthoryear{{McGaugh}, {Lelli}  \& {Schombert}}{{McGaugh}
  et~al.}{2016}]{McGaugh2016}
{McGaugh} S.~S.,  {Lelli} F.,   {Schombert} J.~M.,  2016, \mn@doi [\prl]
  {10.1103/PhysRevLett.117.201101}, \href
  {https://ui.adsabs.harvard.edu/abs/2016PhRvL.117t1101M} {117, 201101}

\bibitem[\protect\citeauthoryear{{McGaugh}, {Li}, {Lelli}  \&
  {Schombert}}{{McGaugh} et~al.}{2018}]{McGaugh2018}
{McGaugh} S.~S.,  {Li} P.,  {Lelli} F.,   {Schombert} J.~M.,  2018, \mn@doi
  [Nature Astronomy] {10.1038/s41550-018-0615-9}, \href
  {https://ui.adsabs.harvard.edu/abs/2018NatAs...2..924M} {2, 924}

\bibitem[\protect\citeauthoryear{{Milgrom}}{{Milgrom}}{1983}]{Milgrom1983}
{Milgrom} M.,  1983, \mn@doi [\apj] {10.1086/161130}, \href
  {https://ui.adsabs.harvard.edu/abs/1983ApJ...270..365M} {270, 365}

\bibitem[\protect\citeauthoryear{{Milgrom}}{{Milgrom}}{1984}]{Milgrom1984}
{Milgrom} M.,  1984, \mn@doi [\apj] {10.1086/162716}, \href
  {https://ui.adsabs.harvard.edu/abs/1984ApJ...287..571M} {287, 571}

\bibitem[\protect\citeauthoryear{{Milgrom}}{{Milgrom}}{2008}]{Milgrom2008}
{Milgrom} M.,  2008, \mn@doi [\nar] {10.1016/j.newar.2008.03.023}, \href
  {https://ui.adsabs.harvard.edu/abs/2008NewAR..51..906M} {51, 906}

\bibitem[\protect\citeauthoryear{{Milgrom}}{{Milgrom}}{2014}]{Milgrom2014}
{Milgrom} M.,  2014, \mn@doi [\prd] {10.1103/PhysRevD.89.024016}, \href
  {https://ui.adsabs.harvard.edu/abs/2014PhRvD..89b4016M} {89, 024016}

\bibitem[\protect\citeauthoryear{{Milgrom}}{{Milgrom}}{2015}]{Milgrom2015}
{Milgrom} M.,  2015, \mn@doi [\mnras] {10.1093/mnras/stv2202}, \href
  {https://ui.adsabs.harvard.edu/abs/2015MNRAS.454.3810M} {454, 3810}

\bibitem[\protect\citeauthoryear{{Milgrom}}{{Milgrom}}{2018}]{Milgrom2018}
{Milgrom} M.,  2018, \mn@doi [\prd] {10.1103/PhysRevD.98.104036}, \href
  {https://ui.adsabs.harvard.edu/abs/2018PhRvD..98j4036M} {98, 104036}

\bibitem[\protect\citeauthoryear{{Milgrom}}{{Milgrom}}{2019}]{Milgrom2019}
{Milgrom} M.,  2019, \mn@doi [\prd] {10.1103/PhysRevD.99.044041}, \href
  {https://ui.adsabs.harvard.edu/abs/2019PhRvD..99d4041M} {99, 044041}

\bibitem[\protect\citeauthoryear{{Milgrom}}{{Milgrom}}{2020}]{Milgrom20}
{Milgrom} M.,  2020, \mn@doi [Studies in the History and Philosophy of Modern
  Physics] {10.1016/j.shpsb.2020.02.004}, \href
  {https://ui.adsabs.harvard.edu/abs/2020SHPMP..71..170M} {71, 170}

\bibitem[\protect\citeauthoryear{{Milgrom} \& {Sanders}}{{Milgrom} \&
  {Sanders}}{2003}]{MS03}
{Milgrom} M.,  {Sanders} R.~H.,  2003, \mn@doi [\apjl] {10.1086/381138}, \href
  {https://ui.adsabs.harvard.edu/abs/2003ApJ...599L..25M} {599, L25}

\bibitem[\protect\citeauthoryear{{Mo}, {Mao}  \& {White}}{{Mo}
  et~al.}{1998}]{Mo1998}
{Mo} H.~J.,  {Mao} S.,   {White} S. D.~M.,  1998, \mn@doi [\mnras]
  {10.1046/j.1365-8711.1998.01227.x}, \href
  {https://ui.adsabs.harvard.edu/abs/1998MNRAS.295..319M} {295, 319}

\bibitem[\protect\citeauthoryear{{Mulchaey}, {Davis}, {Mushotzky}  \&
  {Burstein}}{{Mulchaey} et~al.}{1996}]{12t}
{Mulchaey} J.~S.,  {Davis} D.~S.,  {Mushotzky} R.~F.,   {Burstein} D.,  1996,
  \mn@doi [\apj] {10.1086/176629}, \href
  {https://ui.adsabs.harvard.edu/abs/1996ApJ...456...80M} {456, 80}

\bibitem[\protect\citeauthoryear{{M{\"u}ller}, {Famaey}  \&
  {Zhao}}{{M{\"u}ller} et~al.}{2019}]{Muller2019}
{M{\"u}ller} O.,  {Famaey} B.,   {Zhao} H.,  2019, \mn@doi [\aap]
  {10.1051/0004-6361/201834914}, \href
  {https://ui.adsabs.harvard.edu/abs/2019A&A...623A..36M} {623, A36}

\bibitem[\protect\citeauthoryear{{Navarro}, {Frenk}  \& {White}}{{Navarro}
  et~al.}{1997}]{Navarro1997}
{Navarro} J.~F.,  {Frenk} C.~S.,   {White} S. D.~M.,  1997, \mn@doi [\apj]
  {10.1086/304888}, \href
  {https://ui.adsabs.harvard.edu/abs/1997ApJ...490..493N} {490, 493}

\bibitem[\protect\citeauthoryear{{Nigoche-Netro}, {Aguerri}, {Lagos},
  {Ruelas-Mayorga}, {S{\'a}nchez}  \& {Machado}}{{Nigoche-Netro}
  et~al.}{2010}]{Nigoche-Netro2010}
{Nigoche-Netro} A.,  {Aguerri} J.~A.~L.,  {Lagos} P.,  {Ruelas-Mayorga} A.,
  {S{\'a}nchez} L.~J.,   {Machado} A.,  2010, \mn@doi [\aap]
  {10.1051/0004-6361/200912719}, \href
  {https://ui.adsabs.harvard.edu/abs/2010A&A...516A..96N} {516, A96}

\bibitem[\protect\citeauthoryear{{Oh} et~al.,}{{Oh} et~al.}{2018}]{Oh2018}
{Oh} S.,  et~al., 2018, \mn@doi [\apjs] {10.3847/1538-4365/aacd47}, \href
  {https://ui.adsabs.harvard.edu/abs/2018ApJS..237...14O} {237, 14}

\bibitem[\protect\citeauthoryear{{Osmond} \& {Ponman}}{{Osmond} \&
  {Ponman}}{2004}]{18t}
{Osmond} J. P.~F.,  {Ponman} T.~J.,  2004, \mn@doi [\mnras]
  {10.1111/j.1365-2966.2004.07742.x}, \href
  {https://ui.adsabs.harvard.edu/abs/2004MNRAS.350.1511O} {350, 1511}

\bibitem[\protect\citeauthoryear{{Papastergis}, {Adams}  \& {van der
  Hulst}}{{Papastergis} et~al.}{2016}]{Papastergis2016}
{Papastergis} E.,  {Adams} E.~A.~K.,   {van der Hulst} J.~M.,  2016, \mn@doi
  [\aap] {10.1051/0004-6361/201628410}, \href
  {https://ui.adsabs.harvard.edu/abs/2016A&A...593A..39P} {593, A39}

\bibitem[\protect\citeauthoryear{{Paranjape} \& {Sheth}}{{Paranjape} \&
  {Sheth}}{2021}]{Paranjape2021}
{Paranjape} A.,  {Sheth} R.~K.,  2021, \mn@doi [\mnras]
  {10.1093/mnras/stab2141}, \href
  {https://ui.adsabs.harvard.edu/abs/2021MNRAS.507..632P} {507, 632}

\bibitem[\protect\citeauthoryear{{Pasini} et~al.,}{{Pasini}
  et~al.}{2020}]{Pasini2020}
{Pasini} T.,  et~al., 2020, \mn@doi [\mnras] {10.1093/mnras/staa2049}, \href
  {https://ui.adsabs.harvard.edu/abs/2020MNRAS.497.2163P} {497, 2163}

\bibitem[\protect\citeauthoryear{{Paturel}, {Petit}, {Prugniel}, {Theureau},
  {Rousseau}, {Brouty}, {Dubois}  \& {Cambr{\'e}sy}}{{Paturel}
  et~al.}{2003}]{Paturel2003}
{Paturel} G.,  {Petit} C.,  {Prugniel} P.,  {Theureau} G.,  {Rousseau} J.,
  {Brouty} M.,  {Dubois} P.,   {Cambr{\'e}sy} L.,  2003, \mn@doi [\aap]
  {10.1051/0004-6361:20031411}, \href
  {https://ui.adsabs.harvard.edu/abs/2003A&A...412...45P} {412, 45}

\bibitem[\protect\citeauthoryear{{Peebles}}{{Peebles}}{1996}]{13t}
{Peebles} P.~J.~E.,  1996, \mn@doi [\apj] {10.1086/178125}, \href
  {https://ui.adsabs.harvard.edu/abs/1996ApJ...473...42P} {473, 42}

\bibitem[\protect\citeauthoryear{{Pradyumna} \& {Desai}}{{Pradyumna} \&
  {Desai}}{2021}]{Pradyumna2021b}
{Pradyumna} S.,  {Desai} S.,  2021, \mn@doi [Physics of the Dark Universe]
  {10.1016/j.dark.2021.100854}, \href
  {https://ui.adsabs.harvard.edu/abs/2021PDU....3300854P} {33, 100854}

\bibitem[\protect\citeauthoryear{{Pradyumna}, {Gupta}, {Seeram}  \&
  {Desai}}{{Pradyumna} et~al.}{2021}]{Pradyumna2021}
{Pradyumna} S.,  {Gupta} S.,  {Seeram} S.,   {Desai} S.,  2021, \mn@doi
  [Physics of the Dark Universe] {10.1016/j.dark.2020.100765}, \href
  {https://ui.adsabs.harvard.edu/abs/2021PDU....3100765P} {31, 100765}

\bibitem[\protect\citeauthoryear{{Ren}, {Kwa}, {Kaplinghat}  \& {Yu}}{{Ren}
  et~al.}{2019}]{Ren2019}
{Ren} T.,  {Kwa} A.,  {Kaplinghat} M.,   {Yu} H.-B.,  2019, \mn@doi [Physical
  Review X] {10.1103/PhysRevX.9.031020}, \href
  {https://ui.adsabs.harvard.edu/abs/2019PhRvX...9c1020R} {9, 031020}

\bibitem[\protect\citeauthoryear{{Romanowsky}, {Douglas}, {Arnaboldi},
  {Kuijken}, {Merrifield}, {Napolitano}, {Capaccioli}  \&
  {Freeman}}{{Romanowsky} et~al.}{2003}]{Romanowsky03}
{Romanowsky} A.~J.,  {Douglas} N.~G.,  {Arnaboldi} M.,  {Kuijken} K.,
  {Merrifield} M.~R.,  {Napolitano} N.~R.,  {Capaccioli} M.,   {Freeman} K.~C.,
   2003, \mn@doi [Science] {10.1126/science.1087441}, \href
  {https://ui.adsabs.harvard.edu/abs/2003Sci...301.1696R} {301, 1696}

\bibitem[\protect\citeauthoryear{{Sakai}, {Kennicutt}  \& {Moss}}{{Sakai}
  et~al.}{2012}]{Sakai2012}
{Sakai} S.,  {Kennicutt} Robert~C. J.,   {Moss} C.,  2012, \mn@doi [\apjs]
  {10.1088/0067-0049/199/2/36}, \href
  {https://ui.adsabs.harvard.edu/abs/2012ApJS..199...36S} {199, 36}

\bibitem[\protect\citeauthoryear{{Sanders}}{{Sanders}}{1994}]{Sanders1994}
{Sanders} R.~H.,  1994, \aap, \href
  {https://ui.adsabs.harvard.edu/abs/1994A&A...284L..31S} {284, L31}

\bibitem[\protect\citeauthoryear{{Sanders}}{{Sanders}}{1999}]{Sanders1999}
{Sanders} R.~H.,  1999, \mn@doi [\apjl] {10.1086/311865}, \href
  {https://ui.adsabs.harvard.edu/abs/1999ApJ...512L..23S} {512, L23}

\bibitem[\protect\citeauthoryear{{Sanders}}{{Sanders}}{2003}]{Sanders2003}
{Sanders} R.~H.,  2003, \mn@doi [\mnras] {10.1046/j.1365-8711.2003.06596.x},
  \href {https://ui.adsabs.harvard.edu/abs/2003MNRAS.342..901S} {342, 901}

\bibitem[\protect\citeauthoryear{{Sanders}}{{Sanders}}{2010}]{Sanders2010}
{Sanders} R.~H.,  2010, \mn@doi [\mnras] {10.1111/j.1365-2966.2010.16957.x},
  \href {https://ui.adsabs.harvard.edu/abs/2010MNRAS.407.1128S} {407, 1128}

\bibitem[\protect\citeauthoryear{{Sanders}, {Fabian}  \& {Smith}}{{Sanders}
  et~al.}{2011}]{16t}
{Sanders} J.~S.,  {Fabian} A.~C.,   {Smith} R.~K.,  2011, \mn@doi [\mnras]
  {10.1111/j.1365-2966.2010.17561.x}, \href
  {https://ui.adsabs.harvard.edu/abs/2011MNRAS.410.1797S} {410, 1797}

\bibitem[\protect\citeauthoryear{{Skordis} \& {Z{\l}o{\'s}nik}}{{Skordis} \&
  {Z{\l}o{\'s}nik}}{2021}]{SZ21}
{Skordis} C.,  {Z{\l}o{\'s}nik} T.,  2021, \mn@doi [\prl]
  {10.1103/PhysRevLett.127.161302}, \href
  {https://ui.adsabs.harvard.edu/abs/2021PhRvL.127p1302S} {127, 161302}

\bibitem[\protect\citeauthoryear{{Skrutskie} et~al.,}{{Skrutskie}
  et~al.}{2006}]{2mass}
{Skrutskie} M.~F.,  et~al., 2006, \mn@doi [\aj] {10.1086/498708}, \href
  {https://ui.adsabs.harvard.edu/abs/2006AJ....131.1163S} {131, 1163}

\bibitem[\protect\citeauthoryear{{Smith} et~al.,}{{Smith} et~al.}{2004a}]{9t}
{Smith} R.~J.,  et~al., 2004a, \mn@doi [\aj] {10.1086/423915}, \href
  {https://ui.adsabs.harvard.edu/abs/2004AJ....128.1558S} {128, 1558}

\bibitem[\protect\citeauthoryear{{Smith} et~al.,}{{Smith}
  et~al.}{2004b}]{Smith2004}
{Smith} R.~J.,  et~al., 2004b, \mn@doi [\aj] {10.1086/423915}, \href
  {https://ui.adsabs.harvard.edu/abs/2004AJ....128.1558S} {128, 1558}

\bibitem[\protect\citeauthoryear{{Sohn}, {Geller}  \& {Zahid}}{{Sohn}
  et~al.}{2019}]{Sohn2019}
{Sohn} J.,  {Geller} M.~J.,   {Zahid} H.~J.,  2019, \mn@doi [\apj]
  {10.3847/1538-4357/ab2b46}, \href
  {https://ui.adsabs.harvard.edu/abs/2019ApJ...880..142S} {880, 142}

\bibitem[\protect\citeauthoryear{{Tam}, {Umetsu}, {Robertson}  \&
  {McCarthy}}{{Tam} et~al.}{2023}]{Tam23}
{Tam} S.-I.,  {Umetsu} K.,  {Robertson} A.,   {McCarthy} I.~G.,  2023, \mn@doi
  [\apj] {10.3847/1538-4357/ace33f}, \href
  {https://ui.adsabs.harvard.edu/abs/2023ApJ...953..169T} {953, 169}

\bibitem[\protect\citeauthoryear{{Tian} \& {Ko}}{{Tian} \& {Ko}}{2016}]{TK16}
{Tian} Y.,  {Ko} C.-M.,  2016, \mn@doi [\mnras] {10.1093/mnras/stw1697}, \href
  {https://ui.adsabs.harvard.edu/abs/2016MNRAS.462.1092T} {462, 1092}

\bibitem[\protect\citeauthoryear{{Tian}, {Umetsu}, {Ko}, {Donahue}  \&
  {Chiu}}{{Tian} et~al.}{2020}]{Tian20}
{Tian} Y.,  {Umetsu} K.,  {Ko} C.-M.,  {Donahue} M.,   {Chiu} I.~N.,  2020,
  \mn@doi [\apj] {10.3847/1538-4357/ab8e3d}, \href
  {https://ui.adsabs.harvard.edu/abs/2020ApJ...896...70T} {896, 70}

\bibitem[\protect\citeauthoryear{{Tian}, {Yu}, {Li}, {McGaugh}  \& {Ko}}{{Tian}
  et~al.}{2021a}]{Tian21a}
{Tian} Y.,  {Yu} P.-C.,  {Li} P.,  {McGaugh} S.~S.,   {Ko} C.-M.,  2021a,
  \mn@doi [\apj] {10.3847/1538-4357/abe45c}, \href
  {https://ui.adsabs.harvard.edu/abs/2021ApJ...910...56T} {910, 56}

\bibitem[\protect\citeauthoryear{{Tian}, {Cheng}, {McGaugh}, {Ko}  \&
  {Hsu}}{{Tian} et~al.}{2021b}]{Tian21b}
{Tian} Y.,  {Cheng} H.,  {McGaugh} S.~S.,  {Ko} C.-M.,   {Hsu} Y.-H.,  2021b,
  \mn@doi [\apjl] {10.3847/2041-8213/ac1a18}, \href
  {https://ui.adsabs.harvard.edu/abs/2021ApJ...917L..24T} {917, L24}

\bibitem[\protect\citeauthoryear{{Tonry}, {Dressler}, {Blakeslee}, {Ajhar},
  {Fletcher}, {Luppino}, {Metzger}  \& {Moore}}{{Tonry} et~al.}{2001}]{11t}
{Tonry} J.~L.,  {Dressler} A.,  {Blakeslee} J.~P.,  {Ajhar} E.~A.,  {Fletcher}
  A.~B.,  {Luppino} G.~A.,  {Metzger} M.~R.,   {Moore} C.~B.,  2001, \mn@doi
  [\apj] {10.1086/318301}, \href
  {https://ui.adsabs.harvard.edu/abs/2001ApJ...546..681T} {546, 681}

\bibitem[\protect\citeauthoryear{{Tully} \& {Fisher}}{{Tully} \&
  {Fisher}}{1977}]{Tully1977}
{Tully} R.~B.,  {Fisher} J.~R.,  1977, \aap, \href
  {https://ui.adsabs.harvard.edu/abs/1977A&A....54..661T} {54, 661}

\bibitem[\protect\citeauthoryear{{Verheijen}}{{Verheijen}}{2001}]{Verheijen2001}
{Verheijen} M. A.~W.,  2001, \mn@doi [\apj] {10.1086/323887}, \href
  {https://ui.adsabs.harvard.edu/abs/2001ApJ...563..694V} {563, 694}

\bibitem[\protect\citeauthoryear{{Vikhlinin}, {Kravtsov}, {Forman}, {Jones},
  {Markevitch}, {Murray}  \& {Van Speybroeck}}{{Vikhlinin}
  et~al.}{2006}]{Vikhlinin2006}
{Vikhlinin} A.,  {Kravtsov} A.,  {Forman} W.,  {Jones} C.,  {Markevitch} M.,
  {Murray} S.~S.,   {Van Speybroeck} L.,  2006, \mn@doi [\apj]
  {10.1086/500288}, \href
  {https://ui.adsabs.harvard.edu/abs/2006ApJ...640..691V} {640, 691}

\bibitem[\protect\citeauthoryear{{Voit}}{{Voit}}{2005}]{Voit2005}
{Voit} G.~M.,  2005, \mn@doi [Reviews of Modern Physics]
  {10.1103/RevModPhys.77.207}, \href
  {https://ui.adsabs.harvard.edu/abs/2005RvMP...77..207V} {77, 207}

\bibitem[\protect\citeauthoryear{{Von Der Linden}, {Best}, {Kauffmann}  \&
  {White}}{{Von Der Linden} et~al.}{2007}]{VonDerLinden2007}
{Von Der Linden} A.,  {Best} P.~N.,  {Kauffmann} G.,   {White} S. D.~M.,  2007,
  \mn@doi [\mnras] {10.1111/j.1365-2966.2007.11940.x}, \href
  {https://ui.adsabs.harvard.edu/abs/2007MNRAS.379..867V} {379, 867}

\bibitem[\protect\citeauthoryear{{Wechsler} \& {Tinker}}{{Wechsler} \&
  {Tinker}}{2018}]{WT2018}
{Wechsler} R.~H.,  {Tinker} J.~L.,  2018, \mn@doi [\araa]
  {10.1146/annurev-astro-081817-051756}, \href
  {https://ui.adsabs.harvard.edu/abs/2018ARA&A..56..435W} {56, 435}

\bibitem[\protect\citeauthoryear{{Wenger} et~al.,}{{Wenger}
  et~al.}{2000}]{simbad}
{Wenger} M.,  et~al., 2000, \mn@doi [\aaps] {10.1051/aas:2000332}, \href
  {https://ui.adsabs.harvard.edu/abs/2000A&AS..143....9W} {143, 9}

\bibitem[\protect\citeauthoryear{{White}, {Bliton}, {Bhavsar}, {Bornmann},
  {Burns}, {Ledlow}  \& {Loken}}{{White} et~al.}{1999}]{10t}
{White} R.~A.,  {Bliton} M.,  {Bhavsar} S.~P.,  {Bornmann} P.,  {Burns} J.~O.,
  {Ledlow} M.~J.,   {Loken} C.,  1999, \mn@doi [\aj] {10.1086/301103}, \href
  {https://ui.adsabs.harvard.edu/abs/1999AJ....118.2014W} {118, 2014}

\bibitem[\protect\citeauthoryear{{Williams}, {Bureau}  \&
  {Cappellari}}{{Williams} et~al.}{2009}]{Williams2009}
{Williams} M.~J.,  {Bureau} M.,   {Cappellari} M.,  2009, \mn@doi [\mnras]
  {10.1111/j.1365-2966.2009.15582.x}, \href
  {https://ui.adsabs.harvard.edu/abs/2009MNRAS.400.1665W} {400, 1665}

\bibitem[\protect\citeauthoryear{{Wilson} et~al.,}{{Wilson}
  et~al.}{2016}]{Wilson2016}
{Wilson} S.,  et~al., 2016, \mn@doi [\mnras] {10.1093/mnras/stw1947}, \href
  {https://ui.adsabs.harvard.edu/abs/2016MNRAS.463..413W} {463, 413}

\bibitem[\protect\citeauthoryear{{Yamasaki}, {Sato}, {Mitsuishi}  \&
  {Ohashi}}{{Yamasaki} et~al.}{2009}]{yamasaki2009}
{Yamasaki} N.~Y.,  {Sato} K.,  {Mitsuishi} I.,   {Ohashi} T.,  2009, \mn@doi
  [\pasj] {10.1093/pasj/61.sp1.S291}, \href
  {https://ui.adsabs.harvard.edu/abs/2009PASJ...61S.291Y} {61, S291}

\bibitem[\protect\citeauthoryear{{Zabludoff} \& {Mulchaey}}{{Zabludoff} \&
  {Mulchaey}}{1998}]{Zabludoff1998}
{Zabludoff} A.~I.,  {Mulchaey} J.~S.,  1998, \mn@doi [\apj] {10.1086/305355},
  \href {https://ui.adsabs.harvard.edu/abs/1998ApJ...496...39Z} {496, 39}

\bibitem[\protect\citeauthoryear{{Zappacosta}, {Buote}, {Gastaldello},
  {Humphrey}, {Bullock}, {Brighenti}  \& {Mathews}}{{Zappacosta}
  et~al.}{2006}]{Zappacosta2006}
{Zappacosta} L.,  {Buote} D.~A.,  {Gastaldello} F.,  {Humphrey} P.~J.,
  {Bullock} J.,  {Brighenti} F.,   {Mathews} W.,  2006, \mn@doi [\apj]
  {10.1086/505739}, \href
  {https://ui.adsabs.harvard.edu/abs/2006ApJ...650..777Z} {650, 777}

\bibitem[\protect\citeauthoryear{{Zhang}, {Andernach}, {Caretta}, {Reiprich},
  {B{\"o}hringer}, {Puchwein}, {Sijacki}  \& {Girardi}}{{Zhang}
  et~al.}{2011}]{Zhang2011}
{Zhang} Y.~Y.,  {Andernach} H.,  {Caretta} C.~A.,  {Reiprich} T.~H.,
  {B{\"o}hringer} H.,  {Puchwein} E.,  {Sijacki} D.,   {Girardi} M.,  2011,
  \mn@doi [\aap] {10.1051/0004-6361/201015830}, \href
  {https://ui.adsabs.harvard.edu/abs/2011A&A...526A.105Z} {526, A105}

\bibitem[\protect\citeauthoryear{{Zhao} \& {Famaey}}{{Zhao} \&
  {Famaey}}{2012}]{ZF12}
{Zhao} H.,  {Famaey} B.,  2012, \mn@doi [\prd] {10.1103/PhysRevD.86.067301},
  \href {https://ui.adsabs.harvard.edu/abs/2012PhRvD..86f7301Z} {86, 067301}

\bibitem[\protect\citeauthoryear{{den Hartog} \& {Katgert}}{{den Hartog} \&
  {Katgert}}{1996}]{14t}
{den Hartog} R.,  {Katgert} P.,  1996, \mn@doi [\mnras]
  {10.1093/mnras/279.2.349}, \href
  {https://ui.adsabs.harvard.edu/abs/1996MNRAS.279..349D} {279, 349}

\bibitem[\protect\citeauthoryear{{van den Bosch} \& {Dalcanton}}{{van den
  Bosch} \& {Dalcanton}}{2000}]{vandenBosch2000}
{van den Bosch} F.~C.,  {Dalcanton} J.~J.,  2000, \mn@doi [\apj]
  {10.1086/308750}, \href
  {https://ui.adsabs.harvard.edu/abs/2000ApJ...534..146V} {534, 146}

\makeatother
\end{thebibliography}
